\documentclass{nature}
\usepackage{graphicx,psfrag}
\usepackage{mathrsfs}
\usepackage{amsmath,amsfonts,amssymb}
\usepackage{mathtools}
\usepackage{bbold}
\usepackage{multirow}
\usepackage{comment}
\usepackage{units}
\usepackage{enumerate}
\usepackage[normalem]{ulem}
\usepackage{longtable}
\usepackage{gensymb}
\usepackage{adjustbox}
\usepackage{multibib}
\usepackage{enumitem}
\newcites{New}{References}
\usepackage{hyperref}
\usepackage{caption}
\usepackage{subcaption}
\usepackage{braket}
\usepackage{lipsum} 
\usepackage{xargs}
\usepackage[rgb,pdftex,dvipsnames]{xcolor}
\usepackage{bm}
\usepackage{soul}
\usepackage{booktabs}
\usepackage[colorinlistoftodos,prependcaption,textsize=tiny]{todonotes}
\newcommandx{\unsure}[2][1=]{\todo[linecolor=red,backgroundcolor=red!25,bordercolor=red,#1]{#2}}
\newcommandx{\change}[2][1=]{\todo[linecolor=blue,backgroundcolor=blue!25,bordercolor=blue,#1]{#2}}
\newcommandx{\info}[2][1=]{\todo[linecolor=OliveGreen,backgroundcolor=OliveGreen!25,bordercolor=OliveGreen,#1]{#2}}
\newcommandx{\improvement}[2][1=]{\todo[linecolor=Plum,backgroundcolor=Plum!25,bordercolor=Plum,#1]{#2}}
\newcommandx{\thiswillnotshow}[2][1=]{\todo[disable,#1]{#2}}

\newcommand{\beq}{\begin{equation}}
\newcommand{\eeq}{\end{equation}}
\newcommand{\beqa}{\begin{eqnarray}}
\newcommand{\eeqa}{\end{eqnarray}}

\newcommand{\ben}{\begin{displaymath}}
\newcommand{\een}{\end{displaymath}}
\newcommand{\be}{\begin{equation}}
\newcommand{\ee}{\end{equation}}
\newcommand{\bea}{\begin{eqnarray}}
\newcommand{\eea}{\end{eqnarray}}

\newcommand{\cs}{c_s^2}
\newcommand{\nB}{n_{\!B}}
\newcommand{\nsat}{n_0}
\newcommand{\Esym}{E_{\mathrm{sym}}}

\usepackage{lineno}
\newcounter{enumivTemp}

\title{Generative artificial intelligence for reconstructing neutron-star matter}

\author{Julia Yu.~Panteleeva$^{1}$, Herzallah Alharazin$^{1}$, Evgeny Epelbaum$^{1}$}

\begin{document}

\maketitle

\begin{affiliations}
 \item Institut f\"ur Theoretische Physik II, Ruhr-Universit\"at Bochum,
   D-44780 Bochum, Germany.
\end{affiliations}

\begin{abstract}
Neutron-star cores hold the only known matter in the universe that is simultaneously cold and strongly interacting, compressed beyond nuclear density into a state of unknown composition.~The equation of state links stellar masses, radii and tidal deformabilities to this regime~\cite{Tolman:1939jz,Oppenheimer:1939ne}, but recovering this key quantity from sparse observations is an ill-posed inverse problem.~Existing analyses bury a prior in a fixed functional form, unevenly weighting admissible solutions and biasing the result.~We reconstruct the equation of state with a denoising diffusion model~\cite{SohlDickstein:2015,Ho:2020ddpm} that keeps prior, physics and data separate:~it learns an inspectable, physically motivated prior anchored to first-principles nuclear theory~\cite{Drischler:2020hwi}, while perturbative-QCD~\cite{Gorda:2021znl,Komoltsev:2023zor} and astrophysical~\cite{Miller:2019cac,Salmi:2024aum,Choudhury:2024xbk,LIGOScientific:2018hze} constraints are imposed exactly.~Future measurements therefore will update the posterior by reweighting alone, without retraining or resampling.~The inferred radius of $12.6$~km and tidal deformability of $469$ at $1.4$ solar masses reproduce Gaussian-process~\cite{Essick:2019ldf,Legred:2021hdx,Annala:2023cwx} and heavy-ion-informed~\cite{Huth:2021bsp} inferences despite a far broader prior.~We find near-conformal but still stiff matter in the heaviest stars, consistent with a gradual hadron--quark crossover and disfavouring a strong first-order phase transition.~More broadly, coupling a learned prior to exactly enforced physics establishes a template for ill-posed inverse problems where theory and data constrain different regions~\cite{Alharazin:2026wfh}.
\end{abstract}

Neutron stars concentrate more mass than the Sun into a sphere some twenty kilometres across, compressing their cores to several times the density of atomic nuclei.~They are the only known objects in which cold, strongly interacting matter exists in bulk, and thus the only window into a regime of quantum chromodynamics (QCD) beyond the reach of both laboratory experiments and first-principles computation.~What this matter is made of --- nucleons, hyperons, meson condensates or deconfined quarks --- remains a central open problem of the strong interaction.~Multimessenger astronomy has made the question quantitative:~X-ray pulse-profile modelling, gravitational waves from merging neutron stars and radio timing of the heaviest pulsars now measure their masses, radii and tidal deformabilities precisely enough to constrain the interior composition.

A single function connects those observables to the microphysics.~The pressure of strongly interacting matter depends on density, temperature and composition, but in a mature neutron star the last two are fixed:~the matter is effectively cold, its keV-scale thermal energies being negligible compared with Fermi energies of tens to hundreds of MeV, and weak reactions have long since driven it to charge-neutral $\beta$-equilibrium at every density.~What survives is a single relation between pressure $P$ and energy density $\varepsilon$, the equation of state (EOS), and whatever the constituents, gravity feels them only through this curve.~The general-relativistic equations of stellar structure~\cite{Tolman:1939jz,Oppenheimer:1939ne} take the EOS as their sole microphysical input to yield masses, radii and tidal deformabilities, so measuring these can, in principle, determine the thermodynamics of dense matter.

On the theory side, only limited information about the EOS is available.~At low densities $n \lesssim (1.5$--$2)\nsat$ ($\nsat \approx 0.16$~fm$^{-3}$ is the nuclear saturation density), chiral effective field theory ($\chi$EFT) gives controlled predictions with quantified truncation uncertainties~\cite{Drischler:2020hwi}.~At asymptotically high densities $n \gtrsim 40\nsat$, perturbative QCD (pQCD)~\cite{Kurkela:2009gj,Gorda:2021znl} determines the EOS, and thermodynamic consistency propagates this constraint down to neutron-star densities~\cite{Komoltsev:2021jzg,Komoltsev:2023zor}.~Between these anchors lie the cores of all observed stars, where no controlled calculation exists and model predictions for the pressure already differ by an order of magnitude at $n=2\nsat$.~Across this gap, multimessenger astronomy provides the only empirical information on cold matter, sparse and heterogeneous:~mass--radius measurements from X-ray pulse-profile modelling with NICER~\cite{Miller:2019cac,Salmi:2024aum,Choudhury:2024xbk}, the tidal deformability of the merger GW170817~\cite{LIGOScientific:2018hze} and lower bounds on the maximum mass from two-solar-mass pulsars~\cite{Demorest:2010bx,NANOGrav:2019jur}.~What changes across the gap is the shape of $P(\varepsilon)$ itself, read off from its slope, the squared speed of sound $\cs=\mathrm{d}P/\mathrm{d}\varepsilon$, and above all from how this compares with the conformal value $\cs=1/3$ approached from below by deconfined quark matter in the pQCD limit.~Whether the speed of sound exceeds its conformal value, whether the heaviest stars' cores approach the deconfined-like behaviour of quark matter, and whether the two regimes join smoothly or through a first-order transition can be settled only by combining these constraints~\cite{Annala:2019puf,Annala:2023cwx,Fujimoto:2022ohj}.

Bridging the gap requires solving an ill-posed inverse problem:~reconstructing an unknown function from sparse, noisy and heterogeneous data.~The standard remedy imposes a functional form with a few adjustable parameters, from flexible phenomenological parametrizations to forms tied to specific microphysical scenarios~\cite{Read:2008iy,Hebeler:2013nza,Lindblom:2010bb,Tews:2018kmu,Greif:2018njt,Mueller:1996pm,Nambu:1961tp,McLerran:2018hbz,Alford:2013aca,Kurkela:2009gj,Gorda:2021znl}.~No ansatz is innocent:~each restricts the accessible shapes and weights them unevenly, biasing the posterior~\cite{Greif:2018njt}.~Non-parametric methods relax this but do not remove it.~Gaussian-process priors~\cite{Landry:2018prl,Essick:2019ldf,Legred:2021hdx} admit a broad range of shapes, yet their support is set by the kernel and hyperpriors, often calibrated on tabulated-EOS catalogues, and their smoothness suits gradual behaviour better than the sharp kinks and plateaus of a strong first-order transition.~Machine-learning maps from observables to the EOS~\cite{Fujimoto:2017cdo,Morawski:2020izm,Soma:2022vbb,Ventagli:2024xsh} are highly expressive but learn stellar-structure physics and prior assumptions jointly, entangling them in the network weights so that neither can be investigated separately nor updated as new data arrive without retraining.~Deliberately agnostic analyses~\cite{Huth:2021bsp,Pang:2022rzc} sample the sound speed on a density grid with no global form, but the segmentation, interpolation, bounds and node measure are still hand-chosen and shape the posterior wherever the data run out.~No reconstruction is prior-free; what is desirable is a prior that is explicit, physically interpretable, broad enough that no single parametrization drives the conclusions, and informed by {\it ab initio} theory.

Generative machine learning meets these demands:~rather than fitting a fixed ansatz, it learns a prior over an entire space of admissible functions and reconstructs by conditional generation.~Denoising diffusion models~\cite{SohlDickstein:2015,Ho:2020ddpm}, trained to invert a gradual noising process, are particularly well suited because generation can be steered by inpainting~\cite{Lugmayr:2022repaint}:~theory constraints are imposed exactly where they are trustworthy and the model fills in the rest, a strategy recently used to infer hadronic gravitational form factors from limited lattice-QCD input~\cite{Alharazin:2026wfh}.~Here we apply it to the dense-matter EOS.~A single diffusion model, trained on $10^{6}$ synthetic speed-of-sound profiles spanning various physically motivated functional classes and their combinations (Sec.~\nameref{sec:training}), furnishes a prior that encompasses sound-speed peaks, first-order-transition plateaus and near-conformal tails without committing to any of them, and is shown to be sufficiently general (Sec.~\nameref{sec:methods-prior}).~The $\chi$EFT constraints are imposed during generation as an inpainting anchor, while the pQCD and astrophysical likelihoods are applied afterwards by importance reweighting (Secs.~\nameref{sec:condsample-inpaint} and \nameref{sec:condsample-reweight}).~This keeps prior and likelihood cleanly separated, so future measurements will update the posterior by reweighting alone, without retraining.~The result is a posterior for the squared speed of sound $\cs(\nB)$ between $0.5$ and $8\,\nsat$, with quantified uncertainties and no assumed functional form.~Propagated through the equations of stellar structure, it yields mass--radius and tidal-deformability relations, constrains the slope of the nuclear symmetry energy, and informs the inner-core composition of neutron stars.

\section*{Reconstruction and constraining power}

We draw $10^5$ speed-of-sound profiles from the $\chi$EFT-inpainted diffusion prior, of which $89{,}513$ satisfy causality and thermodynamic stability, $0 \le \cs(\nB) \le 1$, at every point of the density grid.~Each surviving profile is integrated to the pressure and energy density and solved for stellar structure, then reweighted by the marginalized pQCD likelihood~\cite{Komoltsev:2023zor}, three NICER mass–radius posteriors~\cite{Miller:2019cac,Salmi:2024aum,Choudhury:2024xbk}, the GW170817 tidal measurement~\cite{LIGOScientific:2018hze} and the heavy-pulsar maximum-mass bound~\cite{NANOGrav:2017wvv}, (see Secs.~\nameref{sec:condsample-reweight} and \nameref{sec:condsample-like}).~The reweighted posterior retains an effective sample size (ESS), see Eq.~\ref{eq:samp-ess}, of $N_{\mathrm{eff}} = 4{,}187$.~All quantities below are posterior medians with $68\%$ credible intervals unless stated otherwise.

\begin{figure}[tb]
  \begin{center}
\includegraphics[width=0.85 \textwidth,keepaspectratio,angle=0,clip]{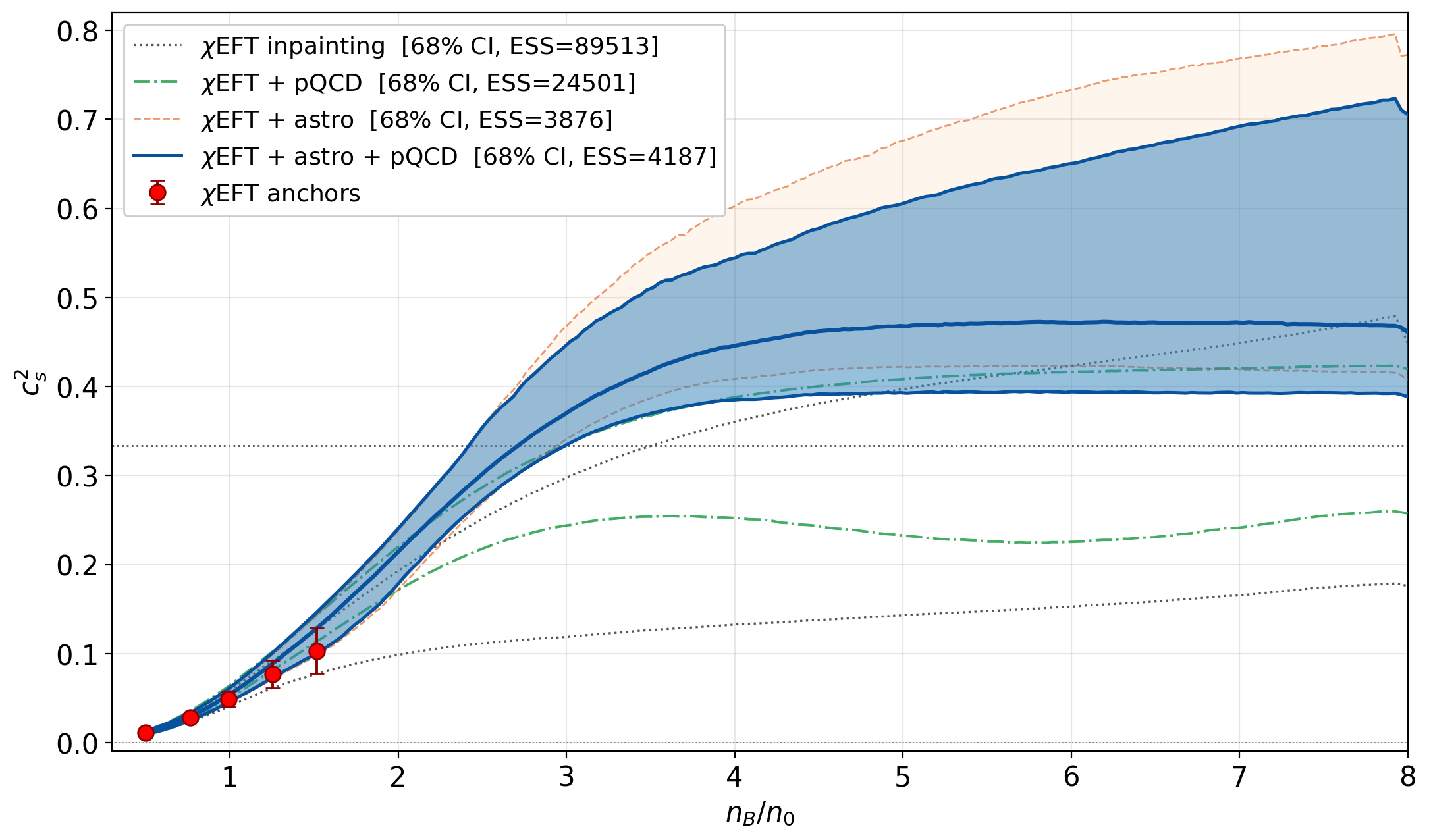}
\end{center}
\linespread{1.3}\selectfont{}
\caption{\textbf{Data ablation of the speed-of-sound posterior.}
The common $\chi$EFT-inpainted prior is reweighted by the pQCD and astrophysical data likelihoods as explained in the legend.~Red points mark the $\chi$EFT anchors.~Bands are pointwise equal-tailed $68\%$ credible intervals.
\label{fig:nested_posteriors}
}
\end{figure}

Reweighting the same inpainted samples with each likelihood separately isolates its constraining power (Fig.~\ref{fig:nested_posteriors}).~With only the $\chi$EFT anchors imposed, the speed of sound above roughly $2\nsat$ is weakly determined.~The pQCD likelihood acts almost exclusively above $\nB \approx 3\nsat$:~it excludes equations of state too soft to connect to the perturbative limit, lifting the high-density band towards $\cs \approx 0.4$ and narrowing it at modest statistical cost ($N_{\mathrm{eff}} = 24{,}501$).~The astrophysical likelihoods instead constrain the density range probed by the observed stars, pushing the band towards the stiffer behaviour demanded by two-solar-mass pulsars and reweighting far more strongly ($N_{\mathrm{eff}} = 3{,}876$).~In the full posterior, the two are complementary:~the astrophysical data fix the canonical-mass region, while pQCD, evaluated at the centre of the maximum-mass star, trims the residual stiff tail back towards the conformal value $\cs = 1/3$ at the highest densities.

\section*{Sound speed and conformality}

We now examine the reconstructed sound speed and its conformality diagnostics.~We use the two publicly released posteriors of
Annala~\textit{et al.}~\cite{Annala:2023cwx}, namely the Gaussian-process ensemble (GP) and the four-segment sound-speed interpolation (C4), as external benchmarks.~Both rest on broad classes of constraints similar to the present work and, therefore, offer the closest available comparison. 

\begin{figure}[tb]
  \begin{center}
\includegraphics[width=\textwidth,keepaspectratio,angle=0,clip]{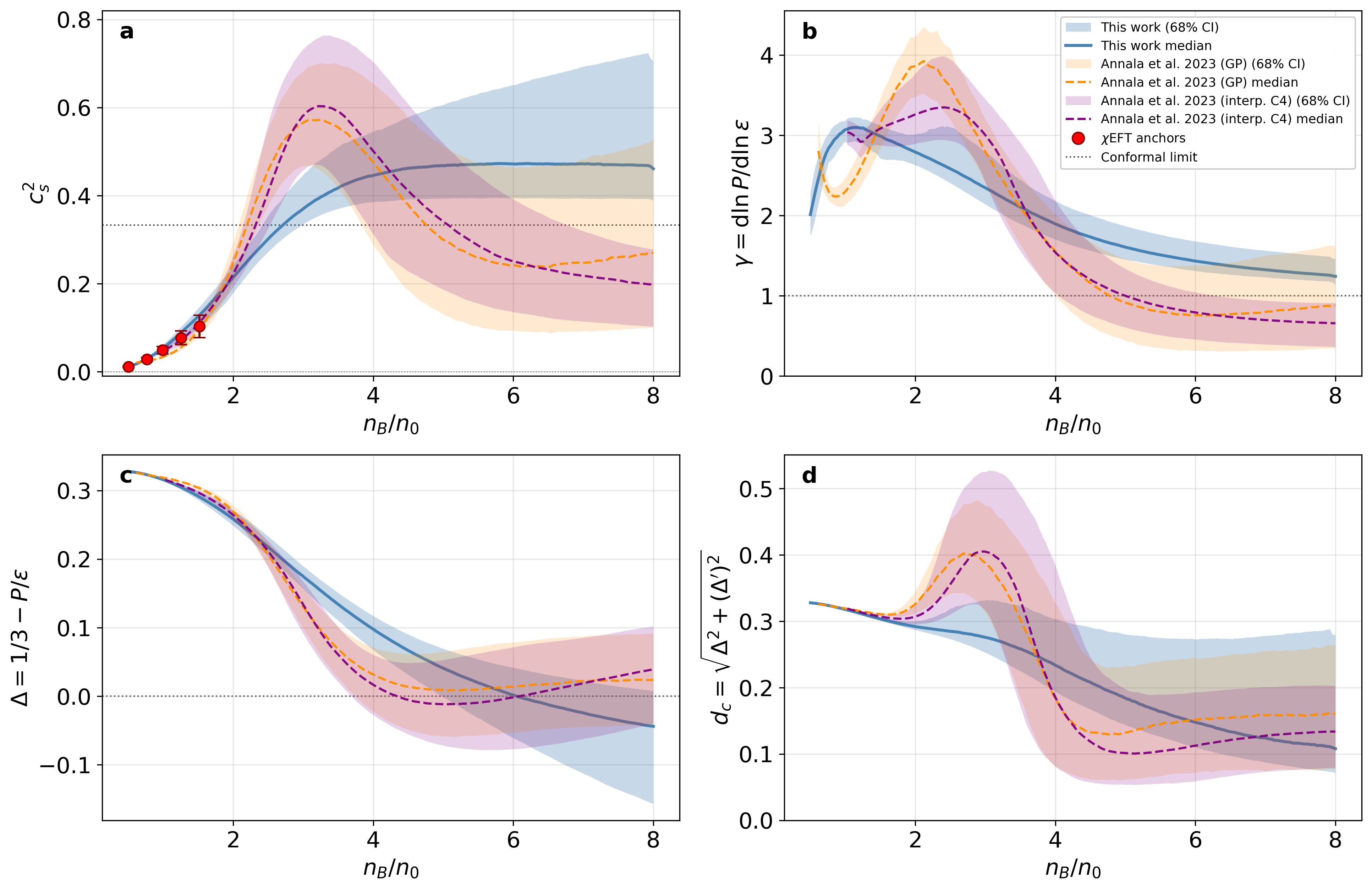}
\end{center}
\linespread{1.3}\selectfont{}
\caption{\textbf{Speed of sound and conformality diagnostics of the reconstructed equation of state.}
All panels are shown versus baryon density.~\textbf{a}, Squared speed of sound $\cs$.~\textbf{b}, Polytropic index $\gamma$.~\textbf{c}, Trace anomaly $\Delta$.~\textbf{d}, Conformal distance $d_c$.~Dotted lines mark the conformal values $\cs = 1/3$ (\textbf{a}), $\gamma = 1$ (\textbf{b}) and $\Delta = 0$ (\textbf{c}).~All results are compared with Gaussian-process (GP) and interpolation (C4) ensembles of Annala~\textit{et al.}~\cite{Annala:2023cwx}.~Shaded bands are pointwise equal-tailed $68\%$ credible intervals and curves their medians.~The $\chi$EFT anchors~\cite{Drischler:2020hwi} are those imposed during generation.
\label{fig:cs2_diagnostics}}
\end{figure}

Figure~\ref{fig:cs2_diagnostics} overlays the three posteriors for the speed of sound and its conformality diagnostics.~All share the low-density $\chi$EFT anchoring but separate above $\nB \approx 2\,\nsat$, where the astrophysical likelihood stiffens the diffusion band relative to its own prior.~The GP and C4 posteriors develop a sharp, well-localized peak in $\cs(\nB)$;~ours does not:~where individual diffusion samples do peak, they peak lower and at higher density, and in much of the posterior $\cs$ is still rising at the stellar centre (see Extended Data Table~\ref{tab:fiducials}), so these features wash out on averaging and the weighted median rises smoothly to a plateau.~Because all three analyses employ comparable $\chi$EFT, astrophysical and perturbative-QCD constraints, this difference primarily reflects the prior:~the diffusion prior spans a broader family of physically motivated equations of state (Sec.~\nameref{sec:training}) than a stationary Gaussian-process kernel or a four-segment parametrization, containing such peaks without imposing them.

The derived diagnostics enter the near-conformal region:~the polytropic index falls through the quark-matter criterion $\gamma < 1.75$~\cite{Annala:2019puf} near $4.5\,\nsat$, the trace anomaly $\Delta = 1/3 - P/\varepsilon$~\cite{Fujimoto:2022ohj} decreases to a median consistent with zero at $n_{\mathrm{TOV}}$, and the median conformal distance $d_c$~\cite{Annala:2023cwx} drops below the near-conformality threshold $0.2$ above $\nB \approx 4.7\,\nsat$.~The speed of sound, by contrast, stays well above its conformal value, $\cs \approx 0.47$ at the centre of the maximum-mass star and at $8\,\nsat$.~Dense matter thus reaches near-conformal bulk thermodynamics while remaining stiff:~$\Delta$ tracks the ratio $P/\varepsilon$ and $\cs$ its slope, so the two need not conformalize together.~Where the GP and interpolation reconstructions have the speed of sound decline after its peak, in step with the thermodynamics, we find that current data do not require that decline.~Indeed, for $45\%$ of the posterior weight $\cs$ has not attained its maximum by the centre of the maximum-mass star (Extended Data Table~\ref{tab:fiducials}).~A turnover must occur, since $\cs$ exceeds $1/3$ at $n_{\mathrm{TOV}}$ and must approach its conformal value asymptotically;~what the data do not fix is whether it lies inside a star.~A fixed functional form places it there by construction.~Brandes~\textit{et al.}~\cite{Brandes:2024wpq} reach a parallel conclusion from the trace anomaly, finding that conformality is approached beyond the densities realized in neutron-star cores.

\section*{Neutron-star structure}

\begin{figure}[tb]
  \begin{center}
\includegraphics[width=\textwidth,keepaspectratio,clip]{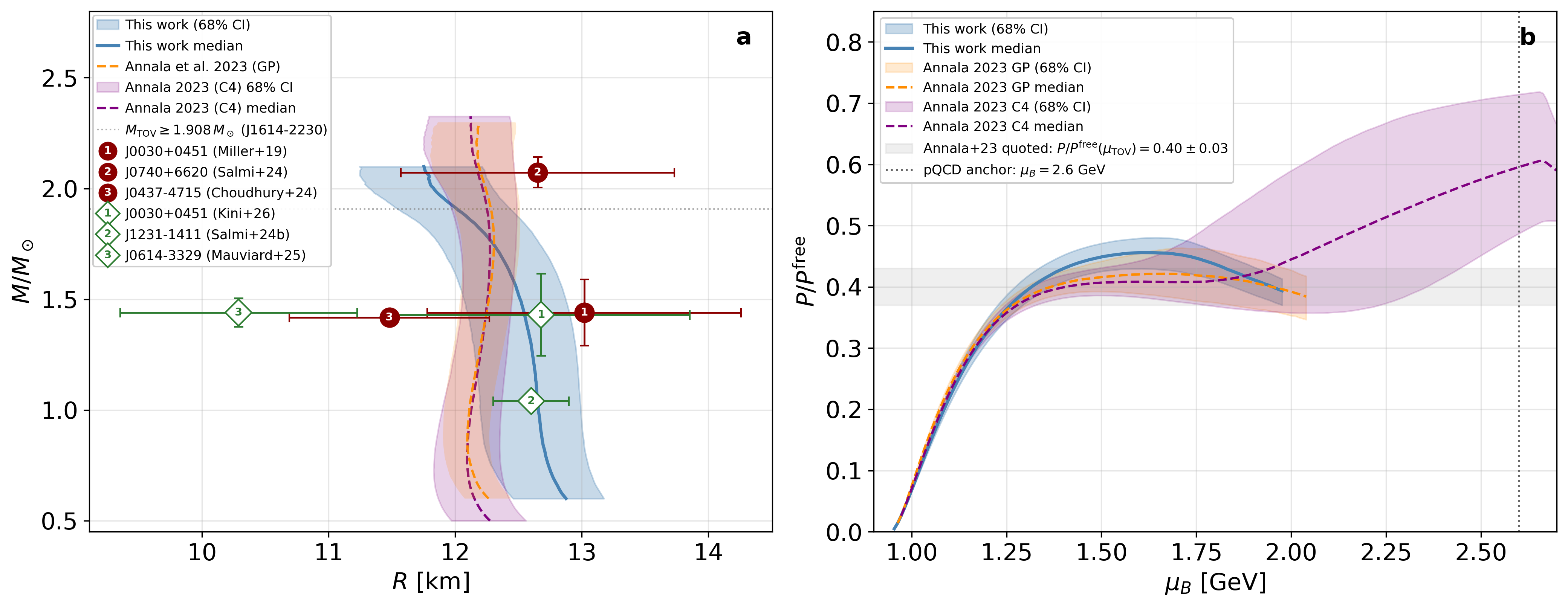}
\end{center}
\linespread{1.3}\selectfont{}
\caption{\textbf{Stellar structure and effective degrees of freedom.}
\textbf{a}, Mass-radius posterior alongside the GP and C4 ensembles of Annala~\textit{et al.}~\cite{Annala:2023cwx}.~Filled circles are the three NICER measurements used as conditioning data~\cite{Miller:2019cac,Salmi:2024aum,Choudhury:2024xbk};~open diamonds are further NICER posteriors withheld from the inference and retained as independent cross-checks~\cite{Kini:2026rjx,Salmi:2024bss,Mauviard:2025dmd}.~The horizontal dotted line is the radio
heavy-pulsar mass bound~\cite{NANOGrav:2017wvv}.~\textbf{b}, Pressure normalized to that of a free gas of massless quarks, $P^{\mathrm{free}} = \mu_B^4/(108\pi^2(\hbar c)^3)$, versus baryon chemical potential;~the grey band reproduces the value at $\mu_{\mathrm{TOV}}$ quoted by Annala~\textit{et al.}~\cite{Annala:2023cwx}.~Bands and error bars are equal-tailed $68\%$ credible intervals and curves their medians, drawn while at least $30\%$ of the respective ensemble weight remains.~C4 is drawn in full, all its samples reaching the perturbative endpoint.
\label{fig:comparison_4panel}}
\end{figure}

Propagating each posterior sample through the stellar-structure and tidal equations (Sec.~\nameref{sec:condsample-like}) yields the mass–radius relation in Fig.~\ref{fig:comparison_4panel}a; the full set of fiducial quantities is collected in Extended Data Table~\ref{tab:fiducials}.

The three NICER measurements used as conditioning data pull in opposite directions and the posterior settles between them (Fig.~\ref{fig:comparison_4panel}a):~PSR~J0030$+$0451~\cite{Miller:2019cac} and PSR~J0740$+$6620~\cite{Salmi:2024aum} favour larger radii, the tightly constrained PSR~J0437$-$4715~\cite{Choudhury:2024xbk} a smaller one, and all three agree with our band within $1.2\sigma$.~Of the three NICER posteriors withheld from the fit, two land on the credible band without ever having informed it~\cite{Kini:2026rjx,Salmi:2024bss}, while PSR~J0614$-$3329~\cite{Mauviard:2025dmd} prefers a markedly smaller radius.~An alternative analysis of PSR~J0437$-$4715~\cite{Miller:2025qfq} shifts $R(1.4\,M_\odot)$ by $0.24$~km (Extended Data Table~\ref{tab:data_variants_a}), so the inferred radius scale does not hinge on a single X-ray analysis (Sec.~\nameref{sec:methods-robust}).

\begin{figure}[tb]
  \begin{center}
\includegraphics[width=\textwidth,keepaspectratio,clip]{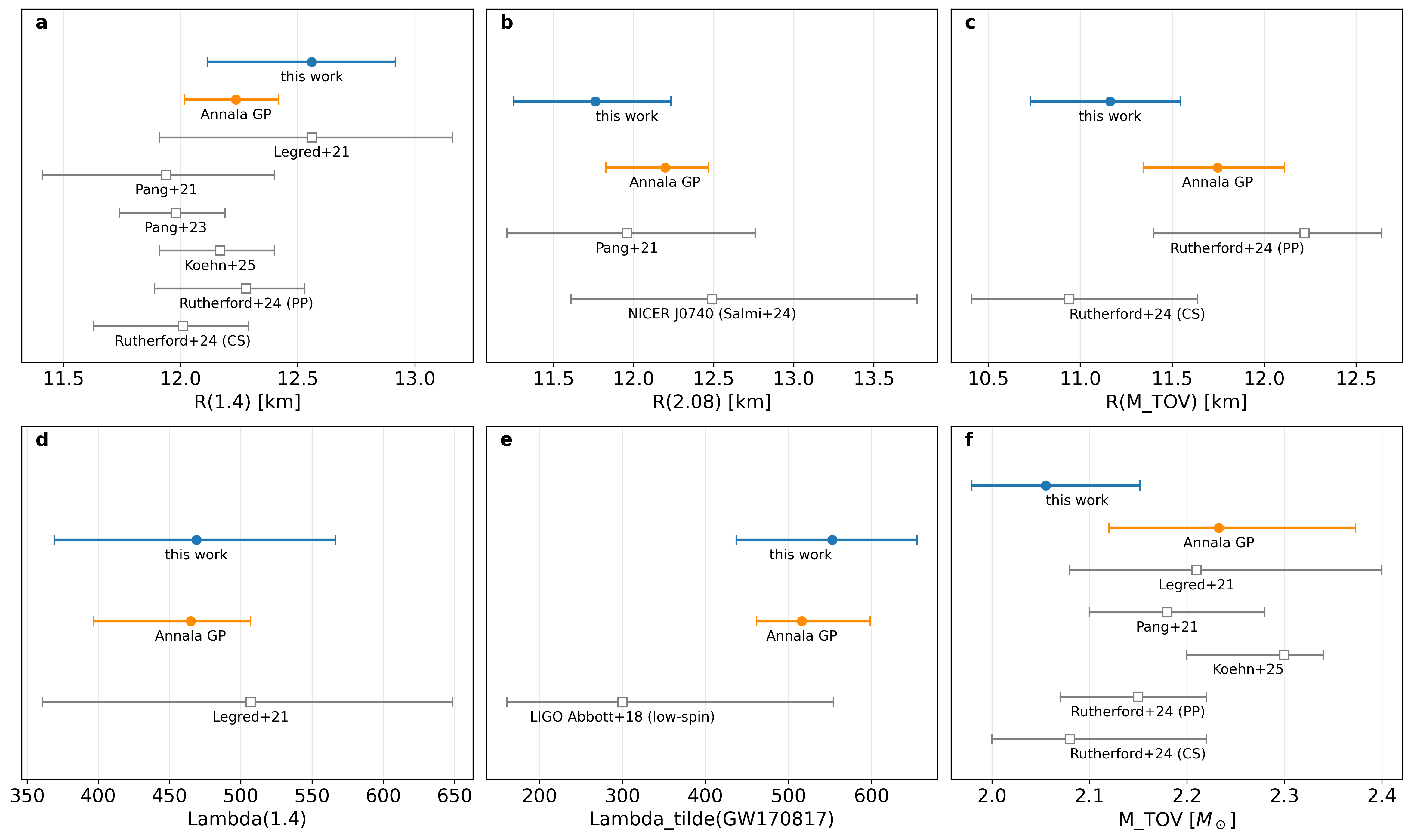}
\end{center}
\linespread{1.3}\selectfont{}
\caption{\textbf{Fiducial neutron-star quantities compared with the literature.}
\textbf{a}--\textbf{c}, Radius at the fixed masses $1.4\,M_\odot$ and $2.08\,M_\odot$,
and at the maximum-mass (TOV) configuration --- the former (\textbf{a}, \textbf{b})
evaluated at a mass shared by all equations of state, the latter (\textbf{c}) at each
equation of state's \emph{own} maximum mass.~\textbf{d}, Tidal deformability at $1.4\,M_\odot$.~\textbf{e}, Mass-weighted binary tidal deformability of GW170817.~\textbf{f}, Maximum mass $M_{\mathrm{TOV}}$.~Filled symbols denote ensemble of the present work and the Annala~2023 GP ensemble~\cite{Annala:2023cwx}.~Open symbols are values as published by Legred~\textit{et al.}~\cite{Legred:2021hdx}, Pang~\textit{et al.}~\cite{Pang:2022rzc}, Koehn~\textit{et al.}~\cite{Koehn:2024set} and Rutherford~\textit{et al.}~\cite{Rutherford:2024srk}, together with the LIGO low-spin posterior~\cite{LIGOScientific:2018hze} in \textbf{e} and the direct NICER radius posterior for PSR~J0740$+$6620~\cite{Salmi:2024aum} in \textbf{b}.~Error bars are
$68\%$ credible intervals;~literature values reported at $90\%$ or $95\%$ credibility have been rescaled assuming approximately Gaussian marginals.
\label{fig:fiducials_forest}}
\end{figure}

Our reconstruction agrees with previous work at the canonical mass and departs from it at high mass (Fig.~\ref{fig:fiducials_forest}):~$R(1.4\,M_\odot)$ and $\Lambda(1.4\,M_\odot)$ lie within the range of Legred~\textit{et al.}~\cite{Legred:2021hdx}, Koehn~\textit{et al.}~\cite{Koehn:2024set}, Pang~\textit{et al.}~\cite{Pang:2022rzc}, Rutherford~\textit{et al.}~\cite{Rutherford:2024srk} and the Gaussian-process (GP) reconstruction of Annala~\textit{et al.}~\cite{Annala:2023cwx}, whereas the values at $2.08\,M_\odot$ and the maximum mass fall below all of them.~The mass--radius relation is thus tilted rather than uniformly shifted, most strongly in the tidal sector, while $\tilde\Lambda(\mathrm{GW170817})$ remains consistent with the LIGO low-spin constraint~\cite{LIGOScientific:2018hze}.

The heaviest stable star also probes deconfinement in its core, at the central baryon density $n_{\mathrm{TOV}} = 6.38^{+0.47}_{-0.45}\,\nsat$, where three proposed diagnostics agree:~the trace anomaly is consistent with the conformal limit, $\Delta(\mathrm{TOV}) = -0.01^{+0.04}_{-0.09}$~\cite{Fujimoto:2022ohj};~the polytropic index falls entirely below the quark-matter criterion, $\gamma(\mathrm{TOV}) = 1.39^{+0.20}_{-0.10} < 1.75$~\cite{Annala:2019puf};~and the conformal distance sits below the near-conformality threshold, $d_c(\mathrm{TOV}) = 0.13^{+0.14}_{-0.03}$ against $d_c < 0.2$~\cite{Annala:2023cwx}, with only its upper tail lying above.~Because each indicator is sensitive to a different combination of $\cs$, $P$ and $\varepsilon$, their agreement at the same time is stronger than any criterion alone.

Our median $\Delta(\mathrm{TOV})$ coincides with $\Delta_{\mathrm{TOV}} = -0.01 \pm 0.03$ of Marczenko~\textit{et al.}~\cite{Marczenko:2022jhl}, on a negatively skewed interval that still admits the conjectured bound $\Delta \ge 0$~\cite{Fujimoto:2022ohj} while leaning towards the $\Delta < 0$ strongly favoured by the NICER-updated Brandes~\textit{et al.}~analysis~\cite{Brandes:2024wpq}, leaving the sign of the trace anomaly unresolved.

A complementary deconfinement measure is the pressure normalized to that of a free gas of massless quarks (Fig.~\ref{fig:comparison_4panel}b).~At the TOV-central chemical potential $\mu_{\mathrm{TOV}} = 1.68^{+0.16}_{-0.09}$~GeV we find $P/P^{\mathrm{free}} = 0.45^{+0.03}_{-0.04}$, consistent with the headline value $0.40 \pm 0.03$ of Annala~\textit{et al.}~\cite{Annala:2023cwx};~evaluated at our $\mu_{\mathrm{TOV}}$ their GP ensemble gives $0.42^{+0.04}_{-0.04}$, so about half the offset follows from their heavier maximum-mass star and correspondingly higher $\mu_{\mathrm{TOV}}$~(Sec.~\nameref{sec:methods-diagnostics}).

\section*{Phase transition and symmetry energy}

The smooth approach to the near-conformal region raises the complementary question of whether the data instead support a strong first-order phase transition.~We test for its necessary, but not sufficient signature~\cite{Brandes:2023hma}:~the probability $\mathcal{P}_{\mathrm{PT}}$ that $\cs$ drops below $0.1$ at a density above its in-window peak, evaluated in $\nB \in [1.5,\,6]\,\nsat$ (Sec.~\nameref{sec:methods-diagnostics}).~Comparing the posterior with the prior value on the same $\chi$EFT-anchored, causality-filtered ensemble isolates the information added by the astrophysical and pQCD likelihoods:~$\mathcal{P}_{\mathrm{PT}}$ falls from $0.088$ to $0.014$, and the posterior-to-prior odds ratio, with the signature in the numerator, yields
\begin{equation}
\mathrm{BF}(\mathrm{PT}\text{ vs.\ }\mathrm{no\text{-}PT})
= \frac{\mathcal{P}_{\mathrm{PT}}^{\mathrm{post}} /
(1 - \mathcal{P}_{\mathrm{PT}}^{\mathrm{post}})}
{\mathcal{P}_{\mathrm{PT}}^{\mathrm{prior}} /
(1 - \mathcal{P}_{\mathrm{PT}}^{\mathrm{prior}})}
= 0.14\,,
\label{eq:BF_PT}
\end{equation}
i.e.~odds of about $7\!:\!1$ against the signature -- substantial, but far from decisive.~The data disfavour a strong transition without excluding it.~The ratio is not fixed by how often such shapes appear in the prior:~ensembles with almost identical prior frequencies return Bayes factors differing by a factor of two, and the ensemble with the highest frequency of threshold crossings returns the lowest Bayes factor of all (Sec.~\nameref{sec:methods-diagnostics}).~An independent multimessenger and heavy-ion analysis~\cite{Huth:2021bsp}, built on a different prior and transition criterion, likewise finds a value below unity ($\approx0.42$), so the conclusion does not rest on our prior's shape content.

Beyond these high-density questions, the same reconstruction reaches back to a nuclear empirical parameter measurable in the laboratory.~For each posterior sample we extract the symmetry-energy slope $L$ from the $\beta$-equilibrium EOS (Sec.~\nameref{sec:methods-L}).~The importance-weighted result
\begin{equation}
L = 60.5^{+7.5}_{-7.6}\,(\mathrm{stat}) \pm 3.0\,(\mathrm{sys})~\mathrm{MeV}
\label{eq:L_result}
\end{equation}
is consistent with the inference $L = 48^{+11}_{-13}$~MeV of Koehn~\textit{et al.}~\cite{Koehn:2024set} and lower than the experiment-driven $L = 83.6^{+19.2}_{-16.8}$~MeV of Tsang~\textit{et al.}~\cite{Tsang:2023vhh}, sitting between the two.~The extraction is stable:~the prior ensemble gives $L = 57.7 \pm 8.3$~MeV and the $\chi$EFT anchoring band $L = 59.8 \pm 4.1$~MeV~\cite{Drischler:2020hwi}, and the astrophysical and pQCD data shift the median only marginally above the $\chi$EFT central value.

\section*{Discussion}
We have reconstructed the neutron-star-matter EOS with a generative prior, conditioned on $\chi$EFT by inpainting and on the astrophysical and pQCD data by importance reweighting.~Because no functional form is imposed, the absence of a sharp sound-speed peak at intermediate density and the persistence of a stiff high-density tail reflect the data rather than an ansatz;~and because reweighting acts on raw samples, multimodality survives conditioning, enabling shape-level tests such as the phase-transition Bayes factor.~With the diffusion network the only learned component, prior and likelihood stay explicitly separated, so the credible intervals are free of emulator bias and carry a quantifiable sampling error.~The posterior presented here is not an endpoint but an instrument.~Every future pulse-profile measurement, tidal signal or sharpened pQCD bound refines it by reweighting alone -- without retraining, without resampling.

The structural conclusions are robust to the composition of the prior ensemble:~training ensembles pushed to the boundary of the prior's support leave the fiducial observables intact (Extended Data Tables~\ref{tab:robustness_a} and~\ref{tab:robustness_b}).~What responds is the sound-speed peak, whose height rises from $0.55$ to at most $0.64$, and the phase-transition Bayes factor, which varies between $0.04$ and $0.24$ but never approaches unity --- so neither the absence of a sharp intermediate-density peak nor the case against a strong transition is an artefact of our baseline shape content (Secs.~\nameref{sec:methods-prior} and \nameref{sec:methods-robust}).

The picture that emerges is coherent: an equation of state softer than previous reconstructions where they peak and stiffer where they decline, whose heaviest stable stars harbour near-conformal matter in its bulk thermodynamics, yet whose sound speed stays well above the conformal limit.~Such a stiff-yet-conformal combination is what a gradual hadron--quark crossover would produce --- quarks emerging and coexisting with hadrons through percolation --- rather than a sharp first-order transition, for which the combined data offer little support.~This reading is consistent with our purely thermodynamic diagnostics but cannot be established by them alone.~The peak is not absent but displaced:~for $45\%$ of the posterior weight $\cs$ has not attained its maximum by the centre of the maximum-mass star, so no mass, radius or tidal deformability can locate it.~Post-merger gravitational-wave signals, which probe transiently higher densities and carry the stiffening in their peak frequency~\cite{Huang:2022mqp}, are the nearest observational handle.

The most substantial extension is in scope.~The EOS inferred here is cold, charge-neutral and $\beta$-equilibrated -- a one-dimensional slice of a far richer thermodynamic surface.~Generalizing the prior to temperature- and composition-dependent equations of state, $P(\nB, T, Y_p)$, would carry the same machinery into neutron-star mergers, proto-neutron-star cooling and core-collapse supernovae, where finite-temperature physics dominates and theory-anchored priors remain scarce.~It can also be confronted with microscopic models of hadronic and quark matter through the diagnostics extracted here, though locating it relative to the deconfined regime requires composition-resolved input.~More broadly, a learned, theory-anchored prior bridging sparse data and firm asymptotic limits addresses a class of inverse problems recurring throughout nuclear and hadronic physics~\cite{Alharazin:2026wfh}.

\setcounter{enumivTemp}{\value{enumiv}}

\newpage

\begin{methods}
  \setcounter{figure}{0}
  \setcounter{table}{0}

\renewcommand{\figurename}{Extended Data Figure}
\renewcommand{\tablename}{Extended Data Table}

\section*{Overview of the inference framework}
\label{sec:methods-overview}
We reconstruct the squared speed of sound $\cs(\nB)$ of cold, charge-neutral, $\beta$-equilibrated matter on a uniform grid of $L=200$ points in baryon density $\nB/\nsat\in[0.5,\,8]$, with $\nsat=0.16~\mathrm{fm^{-3}}$.~A denoising diffusion probabilistic model~\cite{SohlDickstein:2015,Ho:2020ddpm}, trained on a large synthetic ensemble of physically motivated profiles (Sec.~\nameref{sec:training}), serves as a non-parametric prior over $\cs(\nB)$.~Conditioning enters in two cleanly separated stages.~The
chiral-effective-field-theory ($\chi$EFT) constraints at low density are imposed \emph{during} generation, as inpainting anchors~\cite{Lugmayr:2022repaint} on the density grid (Sec.~\nameref{sec:condsample-inpaint}).~The perturbative-QCD (pQCD) and multimessenger astrophysical constraints are imposed \emph{after} generation, as a self-normalized importance-reweighting pass over the raw solutions of the stellar-structure and tidal equations (Secs.~\nameref{sec:condsample-like} and \nameref{sec:condsample-reweight}).~The diffusion network is the only learned component and enters purely as a prior, not as a fitted representation of the equation of state nor as an emulator inside the forward model~\cite{Soma:2022vbb}, so that the reported credible intervals carry only quantifiable sampling error.

\section*{Training ensemble of speed-of-sound profiles}
\label{sec:training}
We train the diffusion model on $10^{6}$ synthetic speed-of-sound profiles, sampled on the density grid defined above, that span the manifold of physically plausible EOS.~The training curves are drawn from ten classes of functional forms:~eight rooted in distinct theoretical or phenomenological approaches to the dense-matter EOS, and two built from convex combinations.~The eight base functional forms are summarized below.~The \emph{piecewise polytrope}~\eqref{eq:fam_poly} is a widely used EOS parametrization~\cite{Vuille:1999,Read:2008iy,Hebeler:2013nza,Zdunik:2005kh,Bejger:2005am,Haensel:2004nu}; we use it to specify the stiffness structure of the EOS by decomposing the density range into $N_{\mathrm{seg}}\in\{3,\cdots, 9\}$ polytropic segments.~The \emph{spectral} class~\eqref{eq:fam_spec} expands the adiabatic index as a polynomial of truncation order $K\in\{2,\cdots, 9\}$ in $\ln(P/P_0)$, with $P_0$ the reference pressure at $0.5\,\nsat$ and expansion coefficients~$\gamma_k$; this representation, introduced in ref.~\cite{Lindblom:2010bb}, accurately reproduces a range of realistic neutron-star EOSs.~The \emph{logistic\,$+$\,Gaussian} model~\eqref{eq:fam_log} follows the speed-of-sound parametrization of Greif~\textit{et al.}~\cite{Greif:2018njt}, combining a logistic baseline matched to the low-density $\chi$EFT regime with a Gaussian peak that lifts $\cs$ above the conformal value $1/3$. It generalizes the earlier ansatz of
Tews~\textit{et al.}~\cite{Tews:2018kmu}, replacing their exponential approach to $1/3$ by a logistic.~The \emph{RMF} class~\eqref{eq:fam_rmf} computes $\varepsilon$ and $P$ from the Walecka model with nonlinear scalar and vector self couplings~\cite{Mueller:1996pm,Boguta:1977xi,Fattoyev:2010mx}, fixed by nuclear saturation properties~\cite{Margueron:2017eqc}, including a density-dependent variant centred on a CompOSE-catalogued parametrization~\cite{Typel:1999yq,Typel:2013rza,CompOSECoreTeam:2022ddl,CompactObject:2024}.

The \emph{CSS} (constant-sound-speed) class~\eqref{eq:fam_css} models a first-order hadron--quark transition~\cite{Alford:2013aca,Alford:2015gna,Han:2019bub,Christian:2017jni}:~a density-dependent hadronic speed of sound $c_{s,\mathrm{H}}^{2}(\nB)$, inherited from a parent class~(1--4), is joined at a transition density $n_{\mathrm{trans}}$ through a Maxwell construction to a quark phase of constant speed of sound $c_{s,\mathrm{QM}}^{2}$; the Seidov criterion~\cite{Seidov:1971} is deliberately not enforced, so that both connected and disconnected (twin-star) branch topologies are represented.~The \emph{quarkyonic} class~\eqref{eq:fam_qy} implements the McLerran--Reddy model~\cite{McLerran:2018hbz,Zhao:2020dvu}:~above a transition density~$n_t$, a quark momentum shell below the baryonic Fermi surface drives a rapid rise of $\cs$ above $1/3$ followed by a gradual recovery toward the conformal value.~The \emph{NJL} crossover~\eqref{eq:fam_njl} realizes the ``three-window'' structure of
NJL/quark--meson models~\cite{Nambu:1961tp,Klevansky:1992qe,Fukushima:2020cmk,Kojo:2021wax}, a stiff chiral crossover with $\cs>1/3$ required to support ${\sim}\,2\,M_\odot$ stars~\cite{Bedaque:2014sqa}, as a conformal baseline plus a Gaussian peak.~Finally, the \emph{pQCD-motivated} class~\eqref{eq:fam_pqcd} enforces the asymptotic approach $\cs\to1/3-\delta_\infty$~\cite{Kurkela:2009gj,Gorda:2021znl}, complementing the quarkyonic and NJL forms in approaching the conformal limit from below.

\begin{subequations}
\label{eq:eos_families}
\begin{align}
&\text{Polytrope:}\quad \cs = \Gamma_i\,\frac{P}{\varepsilon+P}\,, \qquad \Gamma_i=\text{const.\ on segment }i\,,
\label{eq:fam_poly}\\[6pt]
&\text{Spectral:}\quad \cs = \Gamma(P)\,\frac{P}{\varepsilon+P}\,, \qquad \ln\Gamma(P)=\sum_{k=0}^{K}\gamma_k\,\bigl[\ln(P/P_0)\bigr]^{k}\,,
\label{eq:fam_spec}\\[6pt]
&\text{Logistic + Gauss:}\quad \cs(\varepsilon)=\frac{a_1}{1+e^{-a_2(\varepsilon/\varepsilon_0-a_3)}} +a_4\exp\!\biggl[-\frac{(\varepsilon/\varepsilon_0-a_5)^2}{2\,a_6^2}\biggr]\,,
\label{eq:fam_log}\\[6pt]
&\text{RMF:}\quad \cs = \frac{dP/d\nB}{d\varepsilon/d\nB}\,, \qquad \{\varepsilon,P\}\ \text{from the }\sigma\text{--}\omega\text{--}\rho\ \text{mean-field model}\,,
\label{eq:fam_rmf}\\[6pt]
&\text{CSS:}\quad \cs(\nB)=
\begin{cases}
c_{s,\mathrm{H}}^{2}(\nB)\,, & \nB<n_{\mathrm{trans}}\,,\\[2pt]
0\,, & n_{\mathrm{trans}}\le \nB\le \nB^{\mathrm{Q}}\,,\\[2pt]
c_{s,\mathrm{QM}}^{2}\,, & \nB>\nB^{\mathrm{Q}}\,,
\end{cases}
\label{eq:fam_css}\\[6pt]
&\text{Quarkyonic:}\quad \cs = \frac{\nB}{\mu_B}\,\frac{d\mu_B}{d\nB}\,, \qquad \mu_B=\frac{d\varepsilon}{d\nB}\,,\quad \varepsilon(\nB)\ \text{from McLerran--Reddy}\,,
\label{eq:fam_qy}\\[6pt]
&\text{NJL:}\quad \cs(x)=\frac{1}{3}\,\frac{x^{\gamma_1}}{1+x^{\gamma_1}} +A\exp\!\biggl[-\frac{(x-\bar n_\chi)^2}{2\,\sigma_\chi^2}\biggr] \frac{x^{\gamma_2}}{1+x^{\gamma_2}}\,,
\label{eq:fam_njl}\\[6pt]
&\text{pQCD:}\quad \cs(x)=\biggl[\frac{1}{3} +A\exp\!\biggl(-\frac{(x-\bar n_p)^2}{2\,\sigma_p^2}\biggr) -\frac{\delta_\infty}{1+(\bar n_\delta/x)^{\alpha}}\biggr] \frac{x^{\gamma}}{1+x^{\gamma}}\,.
\label{eq:fam_pqcd}
\end{align}
\end{subequations}
Here $x\equiv\nB/\nsat$ is the reduced baryon density; $\varepsilon_0\equiv\nsat\,m_N$ is the nucleon rest-mass energy density at saturation, with $m_N$ the average nucleon mass; and $P$, $\varepsilon$ and $\mu_B$ are the pressure, energy density and baryon chemical potential.~The final two classes are convex combinations of the eight base forms:~Class~9 draws pairwise mixtures and Class~10 Dirichlet-weighted mixtures of three to five parents, excluding physically inconsistent quark-onset pairings (Supplementary Information Sec.~\nameref{sec:si-classes}).

For each class, the free parameters are sampled using multiple randomized strategies to ensure broad coverage of the accessible shape space.~All generated curves satisfy the acceptance criteria ($0\le\cs\le1$), ensuring causality and mechanical stability, are free of numerical singularities, and fulfil a minimum-variation requirement ($\mathrm{std}(\cs)>0.01$) that excludes effectively constant profiles.~The low-density reference point for training data is taken from the SLy4 analytical EOS~\cite{Haensel:2004nu} at $\nB^{\mathrm{ref}}=0.5\,\nsat$.~The complete thermodynamic relations, defining equations, parameter ranges and sampling procedures for every class are given in Supplementary Information Secs.~\nameref{sec:si-master} and~\nameref{sec:si-classes}.

\section*{Denoising diffusion probabilistic model}
\label{sec:condsample-ddpm}
We employ a DDPM with $v$-prediction~\cite{Salimans:2022progressive}.~Let $\mathbf{x}_0\in\mathbb{R}^{L}$ denote a $\cs$ profile on the density grid.~The forward (noising) process maps it, at diffusion step
$\tau\in\{1,\dots,T\}$ with $T=1000$, to
\begin{equation}
\mathbf{x}_\tau=\sqrt{\bar\alpha_\tau}\,\mathbf{x}_0
  +\sqrt{1-\bar\alpha_\tau}\,\bm{\epsilon},
\qquad \bm{\epsilon}\sim\mathcal{N}(\mathbf{0},\mathbf{I}),
\label{eq:samp-forward}
\end{equation}
with $\bar\alpha_\tau=\prod\limits_{s=1}^{\tau}\alpha_s$ and $\alpha_s=1-\beta_s$~\cite{SohlDickstein:2015,Ho:2020ddpm}.~The variance schedule is the cosine form of ref.~\cite{Nichol:2021improved},
\begin{equation}
\bar\alpha_\tau=\frac{f(\tau)}{f(0)},\qquad
f(\tau)=\cos^2\!\left(\frac{\tau/T+s}{1+s}\,\frac{\pi}{2}\right),
\quad s=0.008.
\label{eq:samp-cosine}
\end{equation}
The network is trained to predict the velocity
\begin{equation}
\mathbf{v}_\tau=\sqrt{\bar\alpha_\tau}\,\bm{\epsilon}
  -\sqrt{1-\bar\alpha_\tau}\,\mathbf{x}_0.
\label{eq:samp-vpred}
\end{equation}
Given the network output $\hat{\mathbf{v}}_\tau$, the linear relations~(\ref{eq:samp-forward}) and~(\ref{eq:samp-vpred}) invert to
\begin{align}
\hat{\mathbf{x}}_0&=\sqrt{\bar\alpha_\tau}\,\mathbf{x}_\tau
  -\sqrt{1-\bar\alpha_\tau}\,\hat{\mathbf{v}}_\tau,
\label{eq:samp-x0}\\
\hat{\bm{\epsilon}}&=\sqrt{1-\bar\alpha_\tau}\,\mathbf{x}_\tau
  +\sqrt{\bar\alpha_\tau}\,\hat{\mathbf{v}}_\tau .
\label{eq:samp-eps}
\end{align}
In addition to the noisy input $\mathbf{x}_\tau$ and diffusion time $\tau$, the denoiser receives a binary mask $\mathbf{m}\in\{0,1\}^{L}$, which is unity at the known (anchor) indices, and a conditioning channel $\mathbf{c}\in\mathbb{R}^{L}$ containing the corresponding anchor values (and zero elsewhere).~The training objective is
\begin{equation}
\mathcal{L}(\theta)=\mathbb{E}_{\mathbf{x}_0,\bm{\epsilon},\tau,\mathbf{m}}
\!\left[\big\|f_\theta(\mathbf{x}_\tau,\tau,\mathbf{m},\mathbf{c})
-\mathbf{v}_\tau\big\|^2\right].
\label{eq:samp-loss}
\end{equation}
The denoiser $f_\theta$ is the one-dimensional ResNet--attention hybrid of ref.~\cite{Alharazin:2026wfh}, operating at fixed resolution $L=200$; architecture and training hyperparameters are given in Supplementary Information Sec.~\nameref{sec:si-arch}.~Sampling starts from $\mathbf{x}_T\sim\mathcal{N}(\mathbf{0},\mathbf{I})$ and iterates the ancestral reverse step~\cite{Ho:2020ddpm}
\begin{equation}
\mathbf{x}_{\tau-1}=\tilde{\bm{\mu}}_\tau
  +\sqrt{\tilde\beta_\tau}\,\mathbf{z},
\qquad \mathbf{z}\sim\mathcal{N}(\mathbf{0},\mathbf{I})
\quad(\tau>1),
\label{eq:samp-reverse}
\end{equation}
with the mean and variance of the Gaussian posterior
$q(\mathbf{x}_{\tau-1}\mid\mathbf{x}_\tau,\hat{\mathbf{x}}_0)$,
\begin{align}
\tilde{\bm{\mu}}_\tau&=
  \frac{\sqrt{\bar\alpha_{\tau-1}}\,\beta_\tau}{1-\bar\alpha_\tau}\,
  \hat{\mathbf{x}}_0
  +\frac{\sqrt{\alpha_\tau}\,(1-\bar\alpha_{\tau-1})}{1-\bar\alpha_\tau}\,
  \mathbf{x}_\tau,
\label{eq:samp-mu}\\[2pt]
\tilde\beta_\tau&=\frac{\beta_\tau\,(1-\bar\alpha_{\tau-1})}
  {1-\bar\alpha_\tau},
\label{eq:samp-betatilde}
\end{align}
and $\hat{\mathbf{x}}_0$ from Eq.~(\ref{eq:samp-x0}); the final step $\tau=1$ is taken deterministically, $\mathbf{x}_0=\tilde{\bm{\mu}}_1$.\footnote{For numerical stability $\hat{\mathbf{x}}_0$ is clipped to $[-5,5]$ in standardized units at every step, a safeguard that leaves typical profiles unaffected.}

\section*{Extraction of $\chi$EFT anchors with correlated uncertainties}
\label{sec:extr-anchor}
The anchors are the Monte-Carlo moments of an ensemble of $N=2\times10^{4}$ realizations of the low-density equation of state,
\begin{equation}
\cs(k)=\frac{1}{N}\sum_{i=1}^{N}\big(\cs\big)^{(i)}(n_k), \qquad C_{kl}=\frac{1}{N-1}\sum_{i=1}^{N}\big[(\cs)^{(i)}(n_k)-\cs(k)\big]\big[(\cs)^{(i)}(n_l)-\cs(l)\big],
\label{eq:anchor_moments}
\end{equation}
evaluated at $n_k/\nsat\in\{0.500,0.764,0.990,1.254,1.518\}$, where $C_{kl}$ is the covariance matrix\footnote{The full covariance is propagated:~neighbouring anchors are correlated at $\rho=0.95$--$0.98$.}.~Each realization is a charge-neutral $npe\mu$ equation of state in $\beta$ equilibrium at $T=0$, for which the squared sound speed is
\begin{equation}
\big(\cs\big)^{(i)}=\frac{dP^{(i)}}{d\varepsilon^{(i)}}=\frac{\nB}{\mu_B^{(i)}}\,\frac{d\mu_B^{(i)}}{d\nB}, \qquad \mu_B^{(i)}=e_N^{(i)}+\nB\,\frac{\partial e_N^{(i)}}{\partial\nB}\bigg|_{x_p}+\mu_e^{(i)}x_p^{(i)},
\label{eq:cs2_muB}
\end{equation}
the composition derivative dropping out of $\mu_B^{(i)}$ because $\partial e_N/\partial x_p+\mu_e$ vanishes at weak equilibrium.~The full $npe\mu$ closure is given in Supplementary Information Sec.~\nameref{sec:si-cha-neutr}, where $\mu_B$ reduces to the PNM and SNM energies $E_N^{(i)}$ and $E_S^{(i)}$ alone:~these two curves are the only stochastic objects in the construction, and every other quantity is a deterministic functional of them.~All density derivatives, including $d\mu_B^{(i)}/d\nB$, are centred differences taken within a single realization, so that their correlation with $E_S^{(i)}$ and $\Esym^{(i)}$ is preserved exactly; each realization is propagated on a grid containing the anchor densities exactly, so that no interpolation enters the anchors and the discretisation error in $\cs$ stays below $0.2\%$, more than an order of magnitude smaller than $\sqrt{C_{kk}}$.

Both curves carry the N$^3$LO truncation error, modelled by the Gaussian process of ref.~\cite{Drischler:2020yad} with the published analysis code~\cite{BUQEYE_NM_repo,BUQEYE_gsum_repo} (interpolation and truncation covariance, $\Lambda_b=600$~MeV, chiral cutoff $\Lambda=500$~MeV).~Because Eq.~(\ref{eq:eN}) mixes $E_N$ and $E_S$, their cross-covariance must be retained; it is recovered from three published blocks by the polarization identity
\begin{equation}
\mathrm{Cov}(E_N,E_S)=\tfrac12\left[\Sigma_{NN}+\Sigma_{SS}-\Sigma_{\Esym\Esym}\right].
\label{eq:polarization}
\end{equation}
Realizations are drawn as $\boldsymbol{\theta}^{(i)}=\bar{\boldsymbol{\mu}}+L\mathbf{z}^{(i)}$, $\mathbf{z}^{(i)}\sim\mathcal{N}(\mathbf{0},\mathbb{1})$, with $LL^{\!\top}$ the assembled joint covariance (symmetrised, its negative eigenvalue rim of $O(10^{-4})~\mathrm{MeV^2}$ clipped); each draw fixes both curves at all densities simultaneously, so that the derivatives inherit the correct joint distribution and the same realizations supply the integration constants $(\varepsilon_{\mathrm{ref}},P_{\mathrm{ref}})$ of Eq.~(\ref{eq:P_from_cs}).~In PNM-only mode the pipeline reproduces the published BUQEYE neutron-matter sound speed~\cite{Drischler:2020hwi,BUQEYE_NM_repo} to $1.0\%$ in the mean and $3.5\%$ in the width (Extended Data Figure~\ref{fig:BEQ-our} and Table~\ref{tab:anchors}).

\section*{Theory conditioning by inpainting}
\label{sec:condsample-inpaint}
The $\chi$EFT input enters as fixed values of $\cs$ at $K=5$ low-density grid points.~The anchors are the charge-neutral $\beta$-equilibrium ($npe\mu$) sound-speed values of the previous section.~Conditioning is the inpainting rule of ref.~\cite{Lugmayr:2022repaint}, applied at every reverse step:~the masked entries are overwritten by a forward-noised copy of $\mathbf{c}$ at the current noise level,
\begin{equation}
\mathbf{x}_{\tau-1}\;\leftarrow\;
\mathbf{m}\odot\!\Big(\sqrt{\bar\alpha_{\tau-1}}\,\mathbf{c}
  +\sqrt{1-\bar\alpha_{\tau-1}}\,\bm{\epsilon}'\Big)
+(\mathbf{1}-\mathbf{m})\odot\mathbf{x}_{\tau-1},
\qquad \bm{\epsilon}'\sim\mathcal{N}(\mathbf{0},\mathbf{I}),
\label{eq:samp-replace}
\end{equation}
with $\odot$ the elementwise product and $\bm{\epsilon}'$ drawn independently at each step; at $\tau=1$ the masked entries are set to the clean values,
$\mathbf{x}_0\leftarrow\mathbf{m}\odot\mathbf{c} +(\mathbf{1}-\mathbf{m})\odot\mathbf{x}_0$.~The $\chi$EFT truncation uncertainty is propagated by giving each sample an independent realization $r$ of the anchor values, held \emph{fixed} along its reverse trajectory,
\begin{equation}
{\cs}^r(n_k) =  \cs(n_k) + \delta^{(k)}_r,
\qquad
\bm{\delta}_n\sim\mathcal{N}(\mathbf{0}, C),
\label{eq:samp-jitter}
\end{equation}
where $C$ is the covariance of the sound-speed anchors discussed in the previous section.

\section*{From the speed of sound to the equation of state}
\label{sec:methods-target}
Because the post-generation constraints act on quantities other than the $\chi$EFT-conditioned $\cs(\nB)$ --- the astrophysical likelihoods on masses, radii and tidal deformabilities, which require $P(\varepsilon)$, and the pQCD likelihood on the termination point $(\nB^{\mathrm{T}},\mu^{\mathrm{T}},P^{\mathrm{T}})$ --- each sampled $\cs(\nB)$ profile is converted back to the full equation of state before reweighting.~Given $\cs(\nB)$, the pressure is reconstructed by integrating
\begin{equation}
P(\nB) = P_{\mathrm{ref}} + \int\limits_{\nB^{\mathrm{ref}}}^{\nB}
dn'\;\cs(n')\,\frac{\varepsilon(n')+P(n')}{n'}\,,
\label{eq:P_from_cs}
\end{equation}
jointly with the first law of thermodynamics, upward from a reference point $(\nB^{\mathrm{ref}},P_{\mathrm{ref}},\varepsilon_{\mathrm{ref}})$ at $\nB^{\mathrm{ref}}=0.5\,\nsat$; the baryon chemical potential entering the pQCD likelihood then follows from the Euler relation.~The reference constants are evaluated on the same joint pure-neutron- and symmetric-matter samples of the BUQEYE N$^3$LO analysis~\cite{Drischler:2020hwi,Drischler:2020yad,BUQEYE_NM_repo,BUQEYE_gsum_repo} from which the sound-speed anchors are constructed, rendering them thermodynamically and statistically consistent with the anchors.

\section*{Stellar structure and the data likelihood}
\label{sec:condsample-like}
The reconstructed equation of state $P(\nB),\varepsilon(\nB)$ is fed to the Tolman--Oppenheimer--Volkoff equations~\cite{Tolman:1939jz,Oppenheimer:1939ne} ($G=c=1$),
\begin{equation}
\frac{dP}{dr}=-\frac{(\varepsilon+P)\,(m+4\pi r^3 P)}{r\,(r-2m)},\qquad
\frac{dm}{dr}=4\pi r^2\varepsilon,
\label{eq:samp-tov}
\end{equation}
integrated outward from $P(0)=P_c$ to the surface $P(R)=0$ over a grid of central pressures $P_c$, yielding $M$, $R$ and the tidal deformability $\Lambda$ as functions of $P_c$.~Stability requires $dM/dP_c\ge0$, and $M_{\max}\equiv\max_{P_c}M(P_c)$ is the maximum mass along the stable branch.~Below $\nB^{\mathrm{ref}}$ the equation of state is matched to the SLy crust in the analytic representation of Haensel \& Potekhin~\cite{Haensel:2004nu}, rescaled so that $\varepsilon(P)$ is continuous at the junction.~Profiles whose stable branch has not terminated at the upper edge of the density grid admit only a lower bound on $M_{\max}$ and are discarded rather than scored ($59{,}547$ of the $89{,}513$ causal samples, $66.5\%$);~under the full likelihood these samples would carry $5.3\%$ of the posterior weight, so all reported posteriors are conditioned on $n_{\mathrm{TOV}} \le 8\,\nsat$.~The dimensionless tidal deformability is
\begin{equation}
\Lambda=\tfrac{2}{3}\,k_2\,C^{-5},\qquad C\equiv M/R,
\label{eq:samp-lambda}
\end{equation}
with $k_2$ the quadrupole tidal Love number~\cite{Hinderer:2007mb}.~The surface value $y\equiv y(R)$ of the metric-perturbation variable obeys
\begin{equation}
r\,y'+y^2+y\,F+r^2 Q=0,
\label{eq:samp-love}
\end{equation}
\begin{equation*}
F=\frac{1-4\pi r^2(\varepsilon-P)}{1-2m/r}, \quad Q=\frac{4\pi\!\left[5\varepsilon+9P+\dfrac{\varepsilon+P}{\cs}\right]}{1-2m/r} -\frac{6}{r^2(1-2m/r)} -\left[\frac{2(m+4\pi r^3 P)}{r(r-2m)}\right]^2.
\end{equation*}
The stable evaluation of $k_2$ follows ref.~\cite{Postnikov:2010yn}.~The thermodynamic, stellar-structure and tidal chain was cross-validated against an independent implementation.

The NICER mass--radius measurements, the GW170817 tidal deformability, the heavy-pulsar maximum-mass bound and the pQCD constraint contribute additive log-likelihoods.~Each NICER pulsar marginalizes its measured mass--radius density $f_p$ over the stable branch under a flat prior on the stellar mass, normalized on $[M_{\min},M_{\max}]$ with $M_{\min}=1\,M_\odot$,
\begin{equation}
\log\mathcal{L}_{\rm NICER}=\sum_p\log\!\left[\frac{1}{M_{\max}-M_{\min}}
\int\limits_{M_{\min}}^{M_{\max}}\! f_p\big(M,R(M)\big)\,dM\right],
\label{eq:samp-nicer}
\end{equation}
where $f_p$ is a product of Gaussians in $M$ and $R$ for pulsars given by summary moments and, for J0740 and J0437,
\begin{equation}
f_p(M,R)=\frac{\mathrm{KDE}_p(M,R)\,\mathcal{L}^{\rm radio}_p(M)}
{\pi_{M,p}(M)\,\pi_{R,p}(R\mid M)} ,
\label{eq:samp-nicerkde}
\end{equation}
the released two-dimensional posterior with the priors of the underlying analysis divided out and the radio mass re-added once as an independent Gaussian.~Here $\pi_{M,p}$ is the radio-timing mass prior~\cite{Fonseca:2021wxt,Reardon:2024rdv} and $\pi_{R,p}$ the X-PSI radius prior, uniform on $[4.4,\,16]$~km and restricted by the compactness condition of the pulse-profile model~\cite{Salmi:2024aum,Choudhury:2024xbk}.~Because $\mathcal{L}^{\rm radio}_p$ and $\pi_{M,p}$ are the same Gaussian, the mass factors cancel and only the radius-prior normalization survives.~The NICER information is incorporated at two stages.~First, all three pulsars (J0030+0451~\cite{Miller:2019cac}, J0740+6620~\cite{Salmi:2024aum} and J0437$-$4715~\cite{Choudhury:2024xbk}) enter as separable Gaussians in $(M,R)$, taken from the moments of the released samples for J0740 and J0437 and from the upper half-widths of the published asymmetric credible interval for J0030.~Secondly, J0740 and J0437 are upgraded to kernel-density likelihoods on the full released posteriors~\cite{Salmi:2024aum_zenodo,Choudhury:2024xbk_zenodo}, with the radio-timing masses of refs.~\cite{Fonseca:2021wxt,Reardon:2024rdv}.~Three further NICER posteriors~\cite{Kini:2026rjx,Salmi:2024bss,Mauviard:2025dmd} are withheld from the fit and used only as consistency checks (Fig.~\ref{fig:comparison_4panel}).

For each equation of state, the tidal deformabilities $\Lambda_1$ and $\Lambda_2$ of the two neutron stars in GW170817 are read from the computed mass--deformability relation at the component masses $(m_1,m_2)=(1.48,\,1.26)\,M_\odot$.~The mass-weighted binary tidal deformability is then~\cite{Favata:2013rwa,Wade:2014vqa}
\begin{equation}
\tilde\Lambda=\frac{16}{13}\,
\frac{(m_1+12m_2)m_1^4\Lambda_1+(m_2+12m_1)m_2^4\Lambda_2}{(m_1+m_2)^5} ,
\label{eq:samp-gwtilde}
\end{equation}
which GW170817 constrains to $\hat{\tilde\Lambda}=300^{+420}_{-230}$ at the $90\%$ highest-posterior-density level under the low-spin prior~\cite{LIGOScientific:2018hze}.~We impose this as an asymmetric Gaussian,
\begin{equation}
\log\mathcal{L}_{\rm GW}
=-\tfrac12\Big(\frac{\tilde\Lambda-\hat{\tilde\Lambda}}{\sigma_\pm}\Big)^2 ,
\qquad
\sigma_\pm=\begin{cases}\sigma_+,&\tilde\Lambda>\hat{\tilde\Lambda},\\[2pt]
\sigma_-,&\tilde\Lambda\le\hat{\tilde\Lambda},\end{cases}
\label{eq:samp-gw}
\end{equation}
with $\sigma_+\approx255$ and $\sigma_-\approx140$ the upper and lower $90\%$ half-widths rescaled to $1\sigma$.

Independently, the EOS must support a maximum mass at least as large as the heaviest well-measured pulsar.~We impose this as a one-sided Gaussian on the configuration maximum mass $M_{\max}=\max_{P_c}M(P_c)$,
\begin{equation}
\log\mathcal{L}_{M_{\max}}
=-\tfrac12\Big(\frac{\min(M_{\max}-M_{\rm low},0)}{\sigma}\Big)^2 ,
\label{eq:samp-mmax}
\end{equation}
with the threshold set by the radio-timing mass of PSR~J1614$-$2230, $M_{\rm low}=1.908\,M_\odot$ and $\sigma=0.016\,M_\odot$~\cite{NANOGrav:2017wvv,Demorest:2010bx}.~PSR~J1614$-$2230 only supplies the threshold and is not a NICER source, avoiding double counting with the J0740 likelihood, whose stable branch already drives $M_{\max}\gtrsim2\,M_\odot$.~The sensitivity of the inference to this threshold is examined in Sec.~\nameref{sec:methods-robust}, where it is replaced by the considerably higher mass inferred for the black-widow pulsar PSR~J0952$-$0607~\cite{Romani:2022jhd}.

Finally, at the termination density $\nB^{\rm T}=\nB^{\rm TOV}$ the reconstructed $(\varepsilon^{\rm T},P^{\rm T})$ give the chemical potential $\mu^{\rm T}=(\varepsilon^{\rm T}+P^{\rm T})/\nB^{\rm T}$, and the termination point is scored by the marginalized pQCD likelihood of Komoltsev~\textit{et al.}~\cite{Komoltsev:2023zor,Komoltsev:2024zenodo},
\begin{equation}
\log\mathcal{L}_{\rm pQCD}=\log\mathcal{L}^{\rm marg}\big(\nB^{\rm T},\mu^{\rm T},P^{\rm T}\big),
\label{eq:samp-pqcd}
\end{equation}
which grades how readily a causal, stable, thermodynamically consistent extension connects it to the pQCD anchor at $\mu_B^{\rm H}=2.6$~GeV, returning $-\infty$ when none exists.

\section*{Data conditioning by importance reweighting}
\label{sec:condsample-reweight}
Let $\{\mathbf{x}^{(n)}\}_{n=1}^{N}$ be the conditioned profiles passing the pointwise causality and stability filter $0\le\cs(\nB)\le1$.~For each, the equation of state is reconstructed and solved, and the total log-likelihood factorizes over the independent data
\begin{equation}
\log\mathcal{L}(\mathbf{x}^{(n)})
=\log\mathcal{L}_{\mathrm{NICER}}
+\log\mathcal{L}_{\mathrm{GW}}
+\log\mathcal{L}_{M_{\max}}
+\log\mathcal{L}_{\mathrm{pQCD}} .
\label{eq:samp-loglike}
\end{equation}
Treating the $\chi$EFT-conditioned prior $p_\theta$ as the proposal, the posterior $\pi(\mathbf{x})\propto p_\theta(\mathbf{x})\,\mathcal{L}(\mathbf{x})$ is represented by the self-normalized importance weights $w_n$~\cite{Landry:2018prl,Mroczek:2023eff}
\begin{equation}
w_n=\frac{\exp\!\big[\log\mathcal{L}(\mathbf{x}^{(n)})\big]}
{\sum_{m=1}^{N}\exp\!\big[\log\mathcal{L}(\mathbf{x}^{(m)})\big]},
\qquad
\mathbb{E}_{\pi}[g]\simeq\sum_{n}w_n\,g(\mathbf{x}^{(n)}),
\label{eq:samp-weights}
\end{equation}
where $g$ is any function of $\mathbf{x}$ and $\mathbb{E}_\pi[g]$ its posterior expectation.~The weights are normalized throughout with the log-sum-exp identity.~Posterior summaries are the weighted moments and quantiles obtained from Eq.~(\ref{eq:samp-weights}).~The pointwise mean and variance are
\begin{equation}
\bar x_j=\sum_{n}w_n\,x^{(n)}_j,
\qquad
\sigma_j^2=\sum_{n}w_n\,\big(x^{(n)}_j-\bar x_j\big)^2 .
\label{eq:samp-wsummary}
\end{equation}
The contraction of the prior under the data is monitored by Kish's effective sample size~\cite{Kish:1965survey}
\begin{equation}
N_{\mathrm{eff}}
=\frac{\big(\sum_n w_n\big)^2}{\sum_n w_n^2}
=\Big(\sum_n w_n^2\Big)^{-1},
\label{eq:samp-ess}
\end{equation}
the last equality holding for normalized weights.~Applying Eqs.~(\ref{eq:samp-weights})--(\ref{eq:samp-ess}) to each likelihood term in isolation yields a per-term $N_{\mathrm{eff}}$, identifying the constraint that contracts the prior most strongly.

\section*{Conformality and deconfinement diagnostics}
\label{sec:methods-diagnostics}
From each reconstructed EOS we compute the polytropic index $\gamma$, the trace anomaly $\Delta$~\cite{Fujimoto:2022ohj}, and the conformal distance $d_c$~\cite{Annala:2023cwx}.~The normalized pressure uses the free three-flavour, three-colour massless-quark reference $P^{\mathrm{free}}(\mu_B)$, for which the ensembles are interpolated onto a common chemical-potential grid.~Per-sample peak diagnostics ($c_{s,\max}^2$ and its location $n_{\mathrm{peak}}$) are the maximum of $\cs$ and its argument over the density grid.

As a necessary but not sufficient signature of a strong first-order transition~\cite{Brandes:2023hma}, we evaluate the probability $\mathcal{P}_{\mathrm{PT}}$ that $\cs$ drops below $0.1$ at some density above its in-window peak, with the peak and the search restricted to $\nB\in[1.5,\,6]\,\nsat$.~The associated Bayes factor is the posterior-to-prior odds ratio of Eq.~(\ref{eq:BF_PT}), evaluated on the same $\chi$EFT-anchored, causality-filtered ensemble so that it isolates the information added by the astrophysical and pQCD likelihoods.~Although the full posterior retains $N_{\mathrm{eff}} \approx 4187$, the flagged feature is carried by only $\approx 33$ effective draws; tightening the threshold to $\cs < 0.05$ strengthens it ($\mathrm{BF} = 0.092$, odds of $11\!:\!1$).~Because the Bayes factor is a posterior-to-prior odds ratio, the overall frequency of transition-like shapes in the prior does not determine it and only a qualitatively different distribution of such shapes could shift it.~The augmented-family variants illustrate this (Table~\ref{tab:robustness_a}):~Class~13 ($M_{\mathrm{III}}$), whose causal continuation is monotone by construction (Sec.~\nameref{sec:si-classes}) and adds no genuine first-order-like drops, registers the highest prior frequency of threshold crossings yet the lowest Bayes factor, the shallow crossings produced by it being the most strongly disfavoured by the data.

As external benchmarks we use the two publicly released posteriors of Annala~\textit{et al.}~\cite{Annala:2023cwx}:~the GP and the four-segment sound-speed interpolation (C4)~\cite{Annala:2023cwx_zenodo_GP,Annala:2023cwx_zenodo_C4}.~Each is reduced to the quantities we report using our own diagnostic and stellar-structure code wherever the release does not already provide them; the two releases expose different products, so the $P$--$\varepsilon$ normalization is shared with our posterior for C4 but native to Annala for GP (Supplementary Information Sec.~\nameref{sec:si-external}).

\section*{Symmetry-energy slope}
\label{sec:methods-L}
From every posterior sample we extract the symmetry-energy slope $L\equiv3\nsat\,(\partial\Esym/\partial\nB)|_{\nsat}$ by the $\beta$-equilibrium reconstruction of ref.~\cite{Essick:2021ezp}, modified to use a $\chi$EFT symmetric-matter baseline, to retain muons, and to read $L$ from a weighted parabolic fit.~Using the charge-neutral $npe\mu$ closure of Supplementary Information Sec.~\nameref{sec:si-cha-neutr}, we subtract the symmetric-matter baseline $E_0\equiv E_S$ --- taken from BUQEYE N$^3$LO samples ($\Lambda=500$~MeV)~\cite{Drischler:2020hwi,Drischler:2020yad} held at their posterior mean --- and the equilibrium lepton contribution to isolate $\Esym(\nB)$, which we fit to
\begin{equation}
\Esym(\nB)=J+L\,\chi+\tfrac12 K_{\mathrm{sym}}\,\chi^{2},\qquad \chi=\frac{\nB-\nsat}{3\nsat},
\label{eq:esym_expansion}
\end{equation}
by inverse-variance-weighted least squares.

The value of the slope in Eq.~\ref{eq:L_result} and its $(\mathrm{stat})$ interval are the median and $68\%$ credible interval of $L$ over the importance-weighted posterior, with the symmetric-matter baseline $E_0$ held at its N$^3$LO mean.~The $(\mathrm{sys})$ term is the $\chi$EFT truncation uncertainty of the symmetry energy at $\nsat$, propagated from the correlated BUQEYE PNM and SNM samples with the PNM--SNM correlation retained.~The regulator-cutoff dependence is assessed separately in Sec.~\nameref{sec:methods-robust} and is not folded into this systematic.

\section*{Robustness to the $\chi$EFT anchoring and the astrophysical data}
\label{sec:methods-robust}
The checks in this section probe two sensitivities:~to the $\chi$EFT anchoring applied to the prior at low density, and to the astrophysical data.~All comparisons are made at matched sample size, so differences reflect these choices rather than Monte-Carlo noise.

To test the anchoring, we repeated the full inference with the $\chi$EFT anchors rebuilt at $\Lambda=450$ instead of $500$~MeV.~Under this change the canonical-mass observables agree within their $68\%$ credible intervals, and the high-density sound speed varies by a small fraction of its interval (see Tables~\ref{tab:robustness_a} and~\ref{tab:robustness_b}).~The symmetry-energy slope is the most cutoff-sensitive quantity, shifting from $60.5$ to $66.8$~MeV;~this regulator dependence exceeds the $\pm3.0$~MeV truncation systematic quoted in Eq.~(\ref{eq:L_result}) and is reported separately rather than folded into it.

To test the data, we replaced the PSR~J0437$-$4715 likelihood of Choudhury~\textit{et al.}~\cite{Choudhury:2024xbk} with the independent analysis of Miller~\textit{et al.}~\cite{Miller:2025qfq,Miller_zenodo}, which reports a larger, bimodal radius, and whose own priors -- flat in the inverse compactness $c^{2}R/(GM)$ rather than in the radius -- are divided out in its place.~This shifts $R(1.4\,M_\odot)$ upward by $0.24$~km and $\Lambda(1.4\,M_\odot)$ by $66$, both within their credible intervals, and leaves the maximum-mass configuration and its conformality diagnostics unchanged;~the symmetry-energy slope moves by $4.2$~MeV, comparable to its quoted systematic.~Adding to the maximum-mass bound from PSR~J1614$-$2230 ($M_{\mathrm{TOV}}\gtrsim1.908\pm0.016\,M_\odot$) the far higher value inferred for PSR~J0952$-$0607 ($2.35\pm0.17\,M_\odot$)~\cite{Romani:2022jhd} shifts $M_{\mathrm{TOV}}$ upward by less than its credible interval ($2.06$$\to$$2.13\,M_\odot$) and raises the peak sound speed ($c_{s,\max}^2 = 0.55\to0.66$), leaving the radii, tidal deformabilities and conformality diagnostics unchanged.~These quantities move little because a one-sided maximum-mass bound acts on the low-$M_{\mathrm{TOV}}$ tail rather than on the bulk of the posterior.~All variations remain within the quoted $68\%$ credible intervals (see Tables~\ref{tab:data_variants_a} and \ref{tab:data_variants_b}).

\section*{Robustness of the prior}
\label{sec:methods-prior}
A remaining concern is whether the reconstruction is driven by the network itself rather than by the data and prior.~We therefore assess the generality of the prior directly.~We verified that the baseline model $M_0$ is neither under- nor over-parametrized (Supplementary Information Sec.~\nameref{sec:si-arch}).~Around the convergence plateau (Fig.~\ref{fig:plateau}), the model sits at the small end of the sufficient-capacity range:~reducing the network -- fewer self-attention layers and ResNet blocks -- increases the loss and degrades sample quality, whereas adding capacity leaves both the loss and the downstream $c_s^2$ posterior essentially unchanged.~Systematically varying the training strategies, the parameter ranges of the eight baseline classes and the number of training curves left the posterior within the same region, confirming that the results are controlled by the data and prior, not by the network architecture, parameter ranges, or training-set size.

Furthermore, we probed the prior itself from two complementary directions:~from \emph{outside} the physically motivated space, by adding two free-form families that are not physically motivated a priori -- causal and thermodynamically stable by construction, but encoding no underlying physical description of dense matter -- namely a Gaussian-process sound speed (class~11) and a piecewise-linear interpolation with random nodes (class~12), to test whether the reconstruction is destabilized when the training distribution is broadened beyond physically motivated shapes;~and from \emph{within} that space, by adding a further physically motivated family, built on the nuclear empirical parameters (class~13), to test whether the ten classes already span it.~The free-form classes are stress tests, not candidate priors.~We trained three further models alongside the baseline $M_0$ -- $M_{\mathrm{I}}$ (baseline classes + class~11), $M_{\mathrm{II}}$ (baseline classes + class~12) and $M_{\mathrm{III}}$ (baseline classes + class~13) (Supplementary Information Sec.~\nameref{sec:si-classes}) -- and repeated the entire analysis with each.

The radii, tidal deformabilities, $M_{\mathrm{TOV}}$, the peak sound speed and all three conformality diagnostics at the maximum-mass central density agree within their $68\%$ credible intervals, and no variant supports a strong first-order transition.~The two tests carry different messages.~The free-form classes (classes~11 and~12) populate causal shapes that the baseline model almost never generates.~Yet even trained on this space of not physically motivated curves, the inference is not misled by the added freedom (Tables~\ref{tab:robustness_a} and~\ref{tab:robustness_b}).~The residual sensitivity lies in the sound-speed peak, lowest for the baseline and rising under the stress tests --- still below the Gaussian-process ensemble and of the four-segment interpolation.~The absence of a sharp intermediate-density peak therefore survives the broadening of the training distribution.

We adopt the baseline as fiducial:~each of its ten families is physically motivated a priori, and the class-13 test confirms that they span this space without further additions.~Classes~11 and~12, though not excluded by current data, remain stress tests rather than candidate priors:~by construction, the baseline concentrates its probability mass on sound-speed shapes with a physical realization rather than on all causal functions, and the stress tests quantify the cost of relaxing that choice.

Taken together with the checks of Sec.~\nameref{sec:methods-robust}, these results establish robustness to the \emph{choice} of prior and to bounded changes in the data; they do not imply that the data alone determine the result.~That the reconstruction remains prior-dependent is shown by the effective sample size ($N_{\mathrm{eff}}\approx 4187$, $4.2\%$ of draws), which indicates the likelihood does not overwhelm the prior, and by the $R(1.4\,M_\odot)$ predictive median, which rises from $12.0$ to $12.6$~km after conditioning:~the data pull towards stiffer matter, but only by half a kilometre, and the radius scale inherited from the prior survives.~The reconstructed radius is therefore a joint consequence of the $\chi$EFT-anchored prior and the observations.

\setcounter{enumivTemp}{\value{enumiv}}

\end{methods}
\section*{Data availability}
All datasets generated and/or analyzed during this study are available
from the corresponding author on reasonable request.

\section*{Code availability}
The trained diffusion models and all codes required to generate the training data, train the model, reproduce the sampling procedure, and perform the analysis presented in this work are publicly available at ref.~\cite{GitHubCode:JuliaRepo}.

\begin{addendum}
 \item We are grateful to Jambul Gegelia, Davide Laudicina and Bao-Dong Sun for helpful discussions about the manuscript.~We would also like to thank Yong-Jia Huang for his comments on our manuscript.~We acknowledge the funding from the European Union's Horizon 2020 research and innovation programme under Grant Agreement No.~885150 (ERC AdG NuclearTheory) and from the Federal Ministry of Research, Technology and Space of Germany (BMFTR) through the ErUM-Data project DEMOS.~Parts of the calculations of this work were performed on the HPC cluster Elysium of the Ruhr University Bochum, subsidised by the DFG (INST 213/1055-1).
 \item[Competing Interests] The authors declare that they have no
   competing financial interests.
 \item[Author contributions]
H.A.\ constructed the training ensemble, designed, implemented and trained the diffusion model.~E.E.\ participated in the analysis and interpretation of the results, validated the findings, and guided the methodological development of the study.~J.Yu.P.\ developed the training strategies, the conditioning and importance-reweighting pipeline and participated in the analysis of the results.~All authors jointly conceived the study, discussed every stage of the analysis, and contributed to the writing of the manuscript.~The project was managed by  E.E.~\item[Correspondence] Correspondence and requests for materials should be addressed to J.~Yu.~Panteleeva~(email: panteleevajuly@gmail.com).
\end{addendum}

\clearpage
\setcounter{figure}{0}
\setcounter{table}{0}
\setcounter{equation}{0}
\renewcommand{\thefigure}{S\arabic{figure}}
\renewcommand{\thetable}{S\arabic{table}}
\renewcommand{\theequation}{S\arabic{equation}}
\renewcommand{\figurename}{Supplementary Figure}
\renewcommand{\tablename}{Supplementary Table}

\begin{center}
{\large\bf Supplementary Information}\\[2pt]
{\it Generative artificial intelligence for reconstructing neutron-star matter}
\end{center}

\vspace{1em}

\section*{S1.\quad Master thermodynamic relations and sampling conventions}
\label{sec:si-master}
The fundamental quantity modelled by the diffusion network is the
squared speed of sound,
\begin{equation}
\cs(\nB) \;\equiv\; \frac{dP}{d\varepsilon}\,,
\label{eq:cs_def}
\end{equation}
which connects the pressure~$P$ to the energy density~$\varepsilon$.~Both are related to the energy per particle~$E/A$ by
\begin{align}
P(\nB) &= \nB^2 \, \frac{d(E/A)}{d\nB}\,,
\label{eq:P_from_EA}\\[4pt]
\varepsilon(\nB) &= \nB\bigl[m_N + E/A(\nB)\bigr]\,,
\label{eq:eps_from_EA}
\end{align}
where $m_N = 938.9$~MeV is the average nucleon mass.~Given a speed-of-sound profile $\cs(\nB)$, the pressure is reconstructed by integrating Eq.~(\ref{eq:P_from_cs}) from a reference point
$(\nB^{\mathrm{ref}}, P_{\mathrm{ref}})$ in the known low-density regime.~The first law of thermodynamics at $T=0$ gives
\begin{equation}
\frac{d\varepsilon}{d\nB}
  = \frac{\varepsilon + P}{\nB}\,,
\label{eq:first_law}
\end{equation}
whose inverse is used by classes that parametrize $\cs$ as a function of~$\varepsilon$ rather than~$\nB$.~The adiabatic index
\begin{equation}
\Gamma(P) = \frac{\varepsilon + P}{P}\,\frac{dP}{d\varepsilon} = \frac{\varepsilon + P}{P}\,\cs\,,
\label{eq:Gamma_def}
\end{equation}
enters the spectral representation and relates the baryon density to the pressure via
\begin{equation}
\frac{d\nB}{dP} = \frac{\nB}{P\,\Gamma(P)}\,.
\label{eq:dnB_dP}
\end{equation}
For segments described by a polytropic EOS, with pressure and energy density enforced to be continuous across segment boundaries, the pressure and energy density in the interval $\rho\in[\rho_{i-1},\rho_i]$ satisfy
\begin{equation}
P(\rho) = K_i\,\rho^{\Gamma_i}\,,
\qquad
d\left( \frac{\varepsilon}{\rho} \right) = - P\, d\left( \frac{1}{\rho} \right)\,,
\label{eq:polytropic}
\end{equation}
where $\rho = m_B\,\nB$.

\section*{S2.\quad Charge-neutral $\beta$-equilibrium matter}
\label{sec:si-cha-neutr}
For cold, charge-neutral $npe\mu$ matter in $\beta$ equilibrium, the nucleonic energy per baryon is approximated by the quadratic isospin expansion
\begin{equation}
e_N(\nB,x_p)=(1-x_p)m_n+x_p m_p+E_S(\nB)+\Esym(\nB)\,f^2,\qquad f\equiv 1-2x_p,
\label{eq:eN}
\end{equation}
with $E_S$ and $E_N=E_S+\Esym$ the energy per particle of symmetric nuclear matter (SNM) and pure neutron matter (PNM) excluding rest masses, $\Esym\equiv E_N-E_S$ the symmetry energy, and beyond-quadratic terms neglected~\cite{Somasundaram:2020chb}. The proton fraction follows from weak equilibrium, $\mu_e=-\partial e_N/\partial x_p$, and charge neutrality with leptons as free relativistic Fermi gases,
\begin{equation}
\mu_e=\Delta m+4\Esym f,\qquad x_p\,\nB=\sum_{\ell=e,\mu}\frac{\big[\mu_e^2-m_\ell^2\big]^{3/2}}{3\pi^2(\hbar c)^3}\,\Theta\big(\mu_e-m_\ell\big),
\label{eq:composition}
\end{equation}
with $\Delta m\equiv m_n-m_p$ and $\mu_\mu=\mu_e$; the first relation is linear in $x_p$, and substituting it into the second leaves a single scalar equation for $\mu_e$, giving $x_p\simeq2.5$--$6.2\%$ over the anchor range with muons appearing at $\nB\approx0.141~\mathrm{fm^{-3}}$. The baryon chemical potential then closes on $E_S$ and $\Esym$ alone,
\begin{equation}
\mu_B=m_n+E_S+\nB\frac{dE_S}{d\nB}+\Esym f(2-f)+\nB f^2\frac{d\Esym}{d\nB},
\label{eq:muB_closed}
\end{equation}
from which $P=\nB\mu_B-\varepsilon$ and $\cs=(\nB/\mu_B)\,d\mu_B/d\nB$ follow at $T=0$. This closure underlies both the $\chi$EFT anchors (Methods, Sec.~\nameref{sec:extr-anchor}) and the symmetry-energy slope (Methods, Sec.~\nameref{sec:methods-L}).

\section*{S3.\quad Construction and sampling of the functional classes}
\label{sec:si-classes}
In what follows we give the prior ranges and sampling algorithms for each class introduced in Methods (Sec.~\nameref{sec:training}).~At the end of each sampling process we apply the acceptance criteria $0 \leq \cs \leq 1$ and $\mathrm{std}(\cs) > 0.01$.~To also represent profiles with more than one local maximum, we let about a third of the draws in the two peak-based classes (Classes~3 and~7) carry a second Gaussian peak.~Furthermore, classes~1 and~3 are generated under an additional quota requiring at least $30\%$ of their accepted curves to show a transition-like drop ($\cs<0.05$ at a density above the in-window maximum in $\nB\in[1.5,\,6]\,\nsat$), so that the network sees such shapes often enough to learn them;~without the quota these rates are $6.8\%$ and $14.3\%$.

\subsection*{Class~1: Piecewise polytropes}

Following Read et~al.~\cite{Read:2008iy}, the density range is decomposed into $N_{\mathrm{seg}} \in \{3,\cdots, 9\}$ polytropic segments.~In each segment $\rho \in [\rho_{i-1},\rho_i]$, the EOS takes the
polytropic form~\eqref{eq:polytropic}, with the energy density\footnote{To derive Eq.~(\ref{eq:app_pp_eps}) we applied the first law of thermodynamics with $\lim_{\rho\to 0}\varepsilon/\rho = 1$ after integrating over the energy in Eq.~(\ref{eq:polytropic}).}
\begin{equation}
\varepsilon(\rho) = (1+a_i)\,\rho
  + \frac{K_i}{\Gamma_i - 1}\,\rho^{\Gamma_i}\,.
\label{eq:app_pp_eps}
\end{equation}
Continuity of pressure and energy density at the dividing densities requires
\begin{equation}
K_{i+1} = K_i\,\rho_i^{\,\Gamma_i - \Gamma_{i+1}}\,,
\qquad
a_i = \frac{\varepsilon(\rho_{i-1})}{\rho_{i-1}}
  - \frac{K_i}{\Gamma_i - 1}\,\rho_{i-1}^{\,\Gamma_i - 1} - 1\,.
\label{eq:app_pp_match}
\end{equation}
The speed of sound in each segment is
\begin{equation}
\cs = \frac{\Gamma_i\,P}{\varepsilon + P}\,,
\label{eq:app_pp_cs}
\end{equation}
evaluated directly on the baryon-density grid.
\noindent\textit{Sampling procedure:}~(1)~Pick $N_{\mathrm{seg}} \in \{3,\cdots, 9\}$.~(2)~Sample $\Gamma_i \in [0.5,\,8]$ for each segment.~(3)~Sample $N_{\mathrm{seg}}-1$ dividing densities in
$(0.6,\,7.5)\,\nsat$.~(4)~Fix $K_1 = P(\rho_0)/\rho_0^{\Gamma_1}$ from the crust EOS at~$\rho_0$.~(5)~Propagate $K_{i+1}$ and continuity constants~$a_i$.~(6)~Evaluate $\cs$ at each grid point.

\subsection*{Class~2: Spectral representation}

The energy and baryon densities in the spectral representation are obtained from Eqs.~\eqref{eq:dnB_dP} and~\eqref{eq:Gamma_def}.~The adiabatic index is expanded as in Eq.~\eqref{eq:fam_spec} and the speed of sound is $\cs(P) = \Gamma(P)\,P/[\varepsilon(P)+P]$.
\noindent\textit{Sampling procedure:}~(1)~Pick $K \in \{2,\cdots, 9\}$.~(2)~Sample $\{\gamma_0,\ldots,\gamma_K\}$ from order-dependent prior ranges (narrower with increasing~$k$).~(3)~Set
$P_0 = P_{\mathrm{ref}}$ from the crust EOS at $0.5\,\nsat$.~(4)~Integrate the ODE~\eqref{eq:dnB_dP} upward in~$P$ until $\nB = 8\,\nsat$; discard if not reached.~(5)~Interpolate $\cs$ onto the $\nB$ grid.

\subsection*{Class~3: Speed-of-sound model (logistic\,$+$\,Gaussian)}

The six free parameters defining the speed-of-sound parametrization in Eq.~\eqref{eq:fam_log} are sampled from
\begin{equation}
a_1 \in \bigl[\tfrac{1}{3},1\bigr],\;\;
a_2 \in [0.05, 2.0],\;\;
a_3 \in [1.5, 37],\;\;
a_4 \in [0.1, 1.5],\;\;
a_5 \in [1.5, 12],\;\;
a_6/a_5 \in [0.05, 2]\,,
\label{eq:app_greif_priors}
\end{equation}
motivated by ref.~\cite{Greif:2018njt}.~\noindent\textit{Sampling procedure:} (1)~Sample $(a_1,\ldots,a_6)$ from the prior ranges~\eqref{eq:app_greif_priors}.~(2)~Integrate $dP/d\varepsilon = \cs(\varepsilon)$ and then $d\nB/d\varepsilon = \nB/(\varepsilon+P)$ upward from the reference point; discard if $\nB = 8\,\nsat$ is not reached.~(3)~Interpolate $\cs$ onto the $\nB$ grid.

\subsection*{Class~4: Relativistic mean-field models}

The energy density and pressure in the RMF are given explicitly by Eqs.~(42) and~(43) of ref.~\cite{Mueller:1996pm}.~For cold, $\beta$-equilibrated $npe\mu$ matter we solve the meson field equations on the density grid and take $\cs=(dP/d\nB)/(d\varepsilon/d\nB)$ by centred finite differences.~\textit{Prior ranges:}~$K_0\in[200,300]$~MeV,
$m^*/m_N|_{\nsat}\in[0.55,0.80]$, $J\in[28,36]$~MeV,  $m_\sigma\in[400,600]$~MeV, $\zeta\in[0,0.06]$, $\Lambda_\omega\in[0,0.05]$. 
\textit{Sampling:}~the six parameters above are sampled, and the couplings $(g_\sigma,g_\omega,g_\rho,b,c)$ are fixed by the five saturation conditions ($E_{\rm sat}$, $P=0$,
$K_0$, $m^*$, $J$) at $\nsat$~\cite{Mueller:1996pm,Margueron:2017eqc}, discarding draws with no solution or $L\notin[30,90]$~MeV;~the EOS is then evaluated on the grid and kept if it satisfies the causality and stability criteria.~\textit{Density-dependent variant~\cite{Typel:1999yq}:}~the meson--nucleon couplings $g_i(\nB)$ are smooth functions of density, drawn from priors centred on the
DD-ME2 parametrization~\cite{Lalazissis:2005de} with $\pm30\%$ widths;~the nonlinear self-couplings are switched off and a density-dependent rearrangement self-energy maintains thermodynamic consistency.

\subsection*{Class~5: CSS with phase transition}

The hadronic phase below~$n_{\mathrm{trans}}$ is taken from a parent class $\mathcal{C}\in\{1,2,3,4\}$, supplying $c_{s,\mathrm{H}}^{2}(\nB)$;~above the transition a Maxwell construction ($\cs=0$) joins it to a quark phase of constant speed of sound $c_{s,\mathrm{QM}}^{2}$, giving the composite profile of Eq.~\eqref{eq:fam_css}.~Continuity of the pressure and of the baryon chemical potential across the transition fixes the quark-matter onset density $\nB^{\mathrm{Q}} = n_{\mathrm{trans}}\, (\varepsilon_{\mathrm{trans}} +\Delta\varepsilon+P_{\mathrm{trans}})/(\varepsilon_{\mathrm{trans}}+P_{\mathrm{trans}})$, where $\varepsilon_{\mathrm{trans}}$ and $P_{\mathrm{trans}}$ are the parent values at~$n_{\mathrm{trans}}$ and $\Delta\varepsilon$ is the
latent-heat jump.~The Seidov criterion~\cite{Seidov:1971}, $\Delta\varepsilon_{\mathrm{crit}}/\varepsilon_{\mathrm{trans}} =\tfrac{1}{2}+\tfrac{3}{2}\,P_{\mathrm{trans}}/\varepsilon_{\mathrm{trans}}$, is \emph{not} enforced, so that both connected and disconnected/twin-star branch topologies are represented. 
\noindent\textit{Prior ranges:}~$n_{\mathrm{trans}}/\nsat\in[1.5,7.0]$, $\Delta\varepsilon/\varepsilon_{\mathrm{trans}}\in[0.01,1.5]$, $c_{s,\mathrm{QM}}^{2}\in[0.05,1.0]$. 
\noindent\textit{Sampling procedure:}~(1)~Draw a parent class~$\mathcal{C}$ with probability~$1/4$ each and draw a $c_{s,\mathrm{H}}^{2}(\nB)$ from its generator.~(2)~Reconstruct $\varepsilon(\nB)$ and $P(\nB)$ from the parent $\cs$ via Eqs.~\eqref{eq:first_law} and~\eqref{eq:P_from_cs}.~(3)~Draw the three CSS parameters from the ranges above and compute $\nB^{\mathrm{Q}}$;~discard if $\nB^{\mathrm{Q}}>8\,\nsat$.~(4)~Construct the composite $\cs(\nB)$ via Eq.~\eqref{eq:fam_css}.

\subsection*{Class~6: Quarkyonic matter (McLerran--Reddy model)}

We adopt the chargeless two-flavour (pure-neutron) quarkyonic construction of McLerran and Reddy~\cite{McLerran:2018hbz} (see also Zhao and Lattimer~\cite{Zhao:2020dvu}).~Above the transition density~$n_t$, neutrons occupy a momentum shell of width $\Delta(k_{Fn})=\Lambda^3/k_{Fn}^2+\kappa\,\Lambda/N_c^2$ [Eq.~(5) of ref.~\cite{McLerran:2018hbz}], with lower edge $k_{0n}=k_{Fn}-\Delta$;~the quark Fermi momenta follow from colour matching and charge neutrality, $k_{Fd}=k_{0n}/N_c$ and $k_{Fu}=k_{Fd}/2^{1/3}$.~The onset condition $k_{0n}(n_t)=0$ fixes $\kappa=N_c^2\!\left(k_{tn}/\Lambda-\Lambda^2/k_{tn}^2\right)$ with $k_{tn}=(3\pi^2 n_t)^{1/3}$ [Eq.~(4) of ref.~\cite{Zhao:2020dvu}], which requires $k_{tn}>\Lambda$. The total baryon density $\nB(k_{Fn})$ [Eq.~(8) of ref.~\cite{McLerran:2018hbz}] is inverted numerically for
$k_{Fn}$ at each grid point (below $n_t$: $k_{0n}=k_{Fd}=k_{Fu}=0$). The energy density, relativistic neutron and quark Fermi integrals, with constituent quark mass $M_Q=M_N/N_c$, plus
the interaction term $n_n V_n(n_n)$ with neutron shell density $n_n=(k_{Fn}^3-k_{0n}^3)/(3\pi^2)$, is evaluated analytically (Eqs.~(6) and~(7) of ref.~\cite{Zhao:2020dvu}).~We generalize the McLerran--Reddy neutron potential to free exponents,
\begin{equation}
V_n(n_n)=a\left(\frac{n_n}{\nsat}\right)^{\!\alpha}
        +b\left(\frac{n_n}{\nsat}\right)^{\!\beta},
\label{eq:qy_Vn}
\end{equation}
and obtain $\mu_B=d\varepsilon/d\nB$, $P=\nB\mu_B-\varepsilon$, and $\cs=(\nB/\mu_B)\,d\mu_B/d\nB$ by centred finite differences on the grid (the last form via the $T=0$ Gibbs--Duhem
relation). \noindent\textit{Prior ranges:}
\begin{eqnarray}
&&\Lambda \in [150,\,600]~\text{MeV},~ n_t/\nsat \in [1.0,\,5.0],~a \in [-40,\,20]~\text{MeV},~ \alpha \in [0.3,\,1.5],
\nonumber
\\
&&b \in [0,\,30]~\text{MeV},~ \beta \in [1.5,\,3.5].
\label{eq:qy_priors}
\end{eqnarray}
\noindent\textit{Sampling procedure:} (1)~Sample the six parameters from~\eqref{eq:qy_priors}; reject if $k_{tn}\le\Lambda$.~(2)~On the 200-point grid, invert $\nB(k_{Fn})$ for $k_{Fn}$ and evaluate $\varepsilon(\nB)$.~(3)~Obtain $\mu_B$, $P$ and $\cs$ by centred finite differences.

\subsection*{Class~7: NJL/quark--meson-inspired crossover model}

The speed of sound is parametrized as in~\eqref{eq:fam_njl} and the corresponding parameters are sampled from the following
\noindent\textit{Prior ranges:}
\begin{equation}
A \in [0.05,\,0.60]\,,\;\;
n_\chi/\nsat \in [1.5,\,5.0]\,,\;\;
\sigma_\chi \in [0.3,\,3.0]\,,\;\;
\gamma_1 \in [3,\,10]\,,\;\;
\gamma_2 \in [2,\,8]\,.
\label{eq:njl_priors}
\end{equation}
\noindent\textit{Sampling procedure:} (1)~Sample the five parameters from the prior ranges~\eqref{eq:njl_priors}.~(2)~Evaluate $\cs(\nB)$ from Eq.~\eqref{eq:fam_njl} at each grid point.

\subsection*{Class~8: pQCD-motivated high-density profiles}

The parameters of the speed of sound parametrized as in~\eqref{eq:fam_pqcd} are sampled from the following
\noindent\textit{Prior ranges:}
\begin{eqnarray}
&& A \in [0.0,\,0.60]\,,\;\;
n_p/\nsat \in [2.0,\,5.0]\,,\;\;
\sigma_p \in [0.3,\,3.0]\,,\;\;
\delta_\infty \in [0.02,\,0.25]\,,\;\;
n_\delta/\nsat \in [1.5,\,4.0]\,,\;\; \nonumber \\
&&\alpha \in [1,\,6]\,,
\gamma \in [2,\,8]\,.
\label{eq:pqcd_ext_priors}
\end{eqnarray}
\noindent\textit{Sampling procedure:}~(1)~Sample the seven parameters from the prior ranges~\eqref{eq:pqcd_ext_priors}.~(2)~Evaluate $\cs(\bar{n})$ from Eq.~\eqref{eq:fam_pqcd} at each grid point.

\subsection*{Classes~9--10: Convex combinations}

The two mixture families that fill the gaps in shape space between the eight base classes are defined in Methods (Sec.~\nameref{sec:training}). Class~9 draws pairwise mixtures $\cs=\lambda\,c_s^{2\,(A)}+(1-\lambda)\,c_s^{2\,(B)}$ with $\lambda\sim\mathcal{U}(0,1)$, excluding the pairs $\{5,6\}$, $\{5,7\}$ and $\{6,7\}$ because Classes~5--7 encode mutually exclusive quark-onset mechanisms in the same density window, whose superposition is unphysical. Class~10 forms Dirichlet-weighted mixtures $\cs=\sum_{m=1}^{M}\lambda_m\,c_s^{2\,(m)}$ of $M\in\{3,4,5\}$ parents, with weights $(\lambda_1,\dots,\lambda_M)\sim\mathrm{Dir}(\alpha,\dots,\alpha)$, concentration $\alpha\sim\mathrm{LogU}(0.1,10)$, and at most one of $\{5,6,7\}$ per draw. In both, the parent profiles are drawn afresh from the generators of Classes~1--8, the mixture is formed pointwise on the grid, and the causality and minimum-variation criteria are applied to the result.

\subsection*{Construction of the additional classes 11--13}

Class 11 is a Gaussian-process family defined on the logit of the squared sound speed,
\begin{equation}
\varphi \sim \mathcal{GP}\!\left(m,\,k\right),\qquad
c_s^{2}(x)=\frac{1}{1+e^{-\varphi(x)}},\qquad x=n_B/n_0,
\label{eq:class11}
\end{equation}
so that $c_s^{2}\in(0,1)$ is causal by construction~\cite{Landry:2018prl,Gorda2023}.~Each curve draws one of twelve equally weighted strategies, which fix the covariance to be either squared-exponential or Mat\'ern-3/2,
\begin{equation}
k_{\rm SE}(x,x')=\eta^{2}\exp\!\left[-\frac{(x-x')^{2}}{2\ell^{2}}\right],
\qquad
k_{3/2}(x,x')=\eta^{2}\left(1+\frac{\sqrt{3}\,|x-x'|}{\ell}\right)
\exp\!\left[-\frac{\sqrt{3}\,|x-x'|}{\ell}\right],
\label{eq:class11kernels}
\end{equation}
and the mean to be either constant, $m=\ln\!\left[\bar c^{2}/(1-\bar c^{2})\right]$, or linear in $x$ between logit endpoints $\ln\!\left[c_{\rm lo}^{2}/(1-c_{\rm lo}^{2})\right]$ and $\ln\!\left[c_{\rm hi}^{2}/(1-c_{\rm hi}^{2})\right]$, so that the mean stiffens with density.~The per-curve hyperparameters are drawn from broad hyperpriors spanning $\eta\in[0.15,3]$, $\ell\in[0.2,6]$ (in units of $n_0$) and $\bar c^{2}\in[0.02,0.70]$, under uniform, log-uniform or beta-shaped draws depending on the strategy;~one strategy fixes the mean to the conformal value $\bar c^{2}=1/3$, and one conditions the process on a soft $\chi$EFT-like anchor $c_s^{2}(0.5\,n_0)\in[0.03,0.15]$ at the lowest grid point.

Class 12 is a non-parametric piecewise-linear family,
\begin{equation}
c_s^{2}(x)=v_{i}+\left(v_{i+1}-v_{i}\right)\frac{x-x_{i}}{x_{i+1}-x_{i}},
\qquad x\in[x_{i},x_{i+1}],
\label{eq:class12}
\end{equation}
in the baryon density $x=n_B/n_0$, with two fixed endpoints at $x_{0}=0.5$ and $x_{M+1}=8$ and $M$ random interior nodes $0.6\le x_{1}<\dots<x_{M}\le 7.5$, $M\in\{1,\dots,8\}$; all $M+2$ node values $v_{i}$ (endpoints included) lie in the full causal interval $[0,1]$ set by thermodynamic stability and causality, and $c_s^{2}$ is obtained by linear interpolation across the whole grid.~As with class 11, each curve draws one of twelve equally weighted strategies, which fix the number of nodes, their placement (uniform or log-uniform in $x$), and the node-value draw: uniform or beta-shaped across all of $[0,1]$, uniform on a stiff or soft sub-range, sorted ascending, or with one interior node pushed high or low, or one endpoint pinned to a soft $\chi$EFT-like or near-conformal value.~The piecewise-linear sound-speed representation with causal sampling follows Annala et al.~\cite{Annala:2019puf}; here it is applied directly in number density, like in the density-space sound-speed ansatz of Tews et al.~\cite{Tews:2018kmu}.

Class 13 is built on the nuclear empirical parameters of ref.~\cite{Margueron:2017eqc}, expanding the energy per nucleon around saturation in $x=(n-n_{\rm sat})/(3n_{\rm sat})$ and the isospin asymmetry $\delta$,
\begin{equation}
\begin{aligned}
e(n,\delta) &= e_{\rm sat}(n)+\delta^{2}\,e_{\rm sym}(n),\\
e_{\rm sat}(n) &= E_{\rm sat}+\tfrac{1}{2}K_{\rm sat}x^{2}
+\tfrac{1}{6}Q_{\rm sat}x^{3}+\tfrac{1}{24}Z_{\rm sat}x^{4},\\
e_{\rm sym}(n) &= E_{\rm sym}+L_{\rm sym}x+\tfrac{1}{2}K_{\rm sym}x^{2}
+\tfrac{1}{6}Q_{\rm sym}x^{3}+\tfrac{1}{24}Z_{\rm sym}x^{4}.
\end{aligned}
\label{eq:class13}
\end{equation}
This is the expansion that defines those parameters, rather than the metamodel built on it, which additionally separates an effective-mass-corrected Fermi-gas kinetic term and multiplies each order by a low-density correction factor.~Closure is by $\beta$-equilibrium and charge neutrality with electrons and muons~\cite{Margueron:2017lup}; $\varepsilon(n)$ and $P(n)$, and hence $c_s^{2}=\mathrm dP/\mathrm d\varepsilon$, follow, with the empirical parameters $(n_{\rm sat},E_{\rm sat},K_{\rm sat},Q_{\rm sat},Z_{\rm sat}, E_{\rm sym},L_{\rm sym},K_{\rm sym},Q_{\rm sym},Z_{\rm sym})$ drawn from broad ranges bracketing the empirical averages and uncertainties of ref.~\cite{Margueron:2017eqc}.~Because the order-4 expansion converges only up to $\sim\!4\,n_{\rm sat}$~\cite{Margueron:2017eqc} and becomes acausal or thermodynamically unstable for the overwhelming majority of parameter draws when continued naively to $8\,n_0$, we use the metamodel only below a matching density $n_m=t\,n_{\rm sat}$ with $t\sim\mathcal U(2,4)$, and impose a causal continuation of $c_s^{2}$ above it, drawn among (i) a constant $c_s^{2}(n)=c_s^{2}(n_m)$, (ii) a linear ramp in $n$ from $c_s^{2}(n_m)$ to an endpoint $c_{\rm end}\sim\mathcal U(0,1)$, and (iii) a linear ramp toward the near-conformal band $c_{\rm end}\sim\mathcal U(0.20,0.45)$; each continuation is a convex interpolation of values in $[0,1]$ and is therefore causal and stable by construction.

\section*{S4.\quad Diffusion-model architecture and training}
\label{sec:si-arch}

The denoiser $f_\theta$ is the one-dimensional ResNet--attention hybrid of ref.~\cite{Alharazin:2026wfh}, operated at the fixed resolution $L=200$ of the density grid with no down- or up-sampling:~all grid points carry comparable physical significance, and pooling would discard information.~The network receives three input channels -- the noisy curve $x_\tau$, the binary anchor mask $m\in\{0,1\}^{L}$, and the condition vector $c\in\mathbb{R}^{L}$ holding the normalized anchor values at the masked positions and zero elsewhere (Methods) -- which is mapped by an initial convolution (kernel size 7) to a constant width of 256 channels.~The core consists of twelve residual blocks arranged in three groups of four, each group followed by a multi-head self-attention layer (four heads)~\cite{Vaswani:2017attention}.~Each residual block applies a GroupNorm--SiLU--convolution sequence twice (kernel size 7, dropout 0.1), with the diffusion timestep injected after the first convolution by feature-wise linear modulation (FiLM)~\cite{Perez:2018film}:~a sinusoidal embedding of $\tau$ is passed through a two-layer MLP of width $4\times256$ and projected to per-channel scale and shift parameters.~The self-attention layers provide a global receptive field across the full grid -- essential for the long-range correlations imposed by thermodynamic consistency and by the interplay of low- and high-density constraints -- at negligible cost for $L=200$ ($200\times200$ attention matrices).~The second convolution of every residual block and the output projection of every attention layer are zero-initialized, so that each block acts as the identity map at the start of training and gradients propagate unimpeded through the deep residual stack.~A final GroupNorm--SiLU--convolution projects back to a single channel.~The resulting network has $\approx 1.94\times10^{7}$ trainable parameters.

To impose the $\chi$EFT anchors at inference, the network must complete a full sound-speed profile from the few low-density values held fixed -- and it must equally be able to generate a profile from nothing.~We therefore train a single network to condition on an arbitrary subset of known grid points, by randomizing sample-by-sample which points are revealed to it (the \emph{mask}).
Conditioning is learned through a random-mask curriculum: for $50\%$ of the training samples the mask exposes $5$--$30$ clean values at uniformly random grid positions; for $30\%$ it exposes $5$--$15$ values clustered in the lowest $30\%$ of the grid, mimicking the low-density $\chi$EFT anchoring applied at inference; and for the remaining $20\%$ the mask is empty, which trains the same network to act as an unconditional prior sampler.

The training set comprises the ten families of Methods with $10^{5}$ accepted curves per class ($10^{6}$ curves in total), generated with fixed per-class seeds; an independently generated validation set of $3\times10^{3}$ curves per class ($3\times10^{4}$ in total), drawn with disjoint seeds, is held out to evaluate the loss during training.~Each grid point is standardized to zero mean and unit variance using statistics of the training split only; the statistics are stored with the model and inverted after sampling to recover physical $c_s^2$ values.

\begin{figure}[t]
  \begin{center}
    \vspace{-1cm}
\includegraphics[width=0.65\textwidth,keepaspectratio,angle=0,clip]{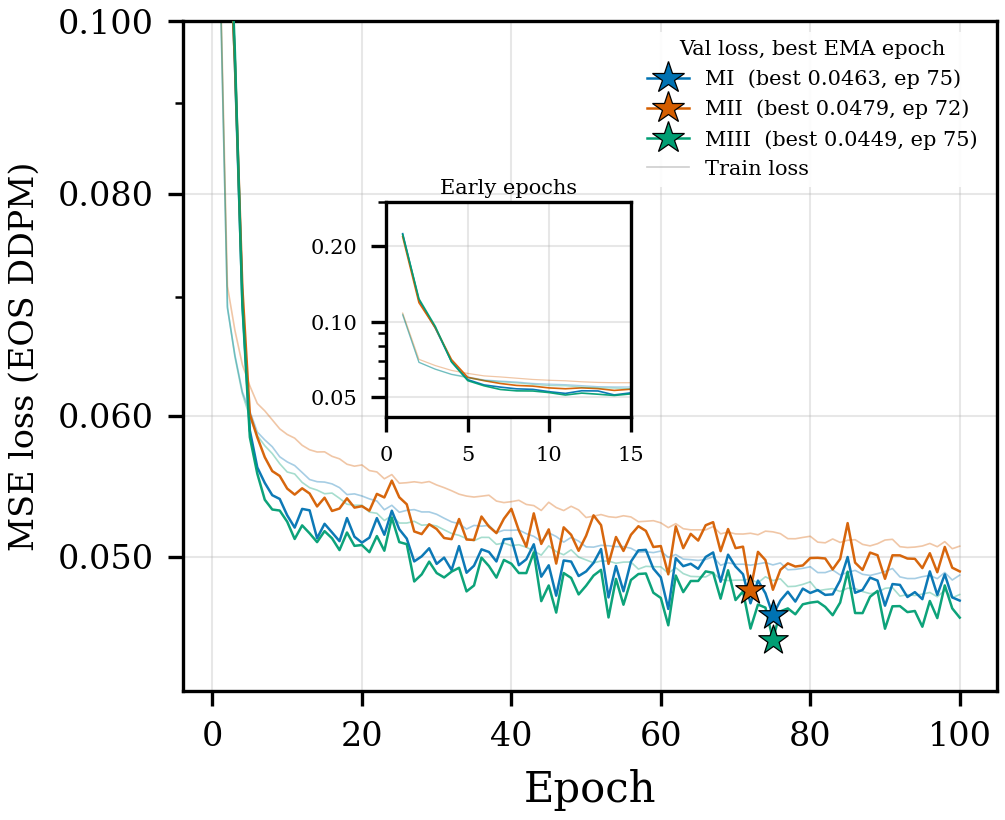}
\end{center}
\linespread{1.3}\selectfont{}
\caption{\textbf{Training dynamics of the augmented-family diffusion models.}~Validation MSE loss over the 100-epoch schedule for $M_{\mathrm{I}}$ (baseline classes $+$ class~11), $M_{\mathrm{II}}$ ($+$ class~12) and $M_{\mathrm{III}}$ ($+$ class~13);~thin curves are the corresponding training losses.~Stars mark the lowest EMA-evaluated validation losses, whose weights define the models used in Extended Data Tables~\ref{tab:robustness_a} and~\ref{tab:robustness_b}.~Inset, the first 15 epochs at full vertical scale.
\label{fig:plateau_variants}
}
\end{figure}

Optimization uses AdamW~\cite{Loshchilov:2019adamw} with learning rate $10^{-4}$, weight decay $10^{-4}$, and batch size $120$, with a linear warm-up over the first two epochs followed by a per-step cosine annealing of the learning rate (100-epoch horizon), gradient-norm clipping at $1.0$; the random seed is fixed for reproducibility.~With $10^{6}$ training curves each epoch comprises $\approx8.3\times10^{3}$ optimizer steps.~An exponential moving average (EMA) of the network weights (decay $0.9999$) is maintained throughout; validation and all sampling use the EMA weights.~The training loss decreases steadily throughout the full $100$-epoch schedule, from $\approx0.052$ at epoch $20$ to $\approx0.047$ at epoch $100$, while the validation loss follows the same slow downward trend, from $\approx0.050$ to $\approx0.046$, with a non-monotonic epoch-to-epoch scatter at the few-percent level that is typical for diffusion-model training.~Beyond the initial $\approx$15 warm-up epochs, the EMA-evaluated validation loss lies at or slightly below the training loss (consistent with no overfitting), and the EMA weights with the lowest validation loss ($0.0439$, epoch $83$) define the baseline model $M_0$.~The augmented-family models $M_{\mathrm{I}}$--$M_{\mathrm{III}}$ are trained with identical architecture, hyperparameters and schedule;~their loss curves follow the same pattern (Extended Data Figure~\ref{fig:plateau_variants}), and the EMA weights at their lowest validation losses define the instances used in Extended Data Tables~\ref{tab:robustness_a} and~\ref{tab:robustness_b}.

\section*{S5.\quad Reprocessing of the external ensembles}
\label{sec:si-external}
The GP and C4 posteriors of Annala~\textit{et al.}~\cite{Annala:2023cwx,Annala:2023cwx_zenodo_GP,Annala:2023cwx_zenodo_C4} are ingested from their public releases and reduced to the quantities of Table~\ref{tab:fiducials}, with each release handled according to what it provides.~For the GP ensemble the released per-sample $(\cs,\varepsilon,P)$ curves and $(M,R,\Lambda,M_{\max})$ arrays are used directly, converted to consistent units and interpolated onto our density grid, and the combined weight is formed from the published likelihood columns;~the conformality diagnostics $(\gamma,\Delta,d_c)$ are computed with our code, the GP release carrying no diagnostic arrays.~For the C4 ensemble only $\cs(\nB)$ is released in re-griddable form, so $\varepsilon$ and $P$ are reconstructed by integrating $\cs$ from the same $\chi$EFT curve used for our own samples, evaluated at $1.03\,\nsat$ (Sec.~\nameref{sec:methods-target}), while the released $(\gamma,\Delta,d_c)$ and $(M,R)$ arrays are used where available.~Consequently the mass--radius and tidal quantities of the two
ensembles are derived from each published stellar-structure solution, whereas in the $P$--$\varepsilon$ and $P/P^{\mathrm{free}}$ plane the C4 band and our posterior share the same $\chi$EFT normalization, applied at $1.03\,\nsat$ for C4 and at $0.5\,\nsat$ for our samples, while the GP band carries Annala's own low-density normalization.~Because that anchoring is itself $\chi$EFT-based, the resulting offset near $0.5\,\nsat$ is small but not removed.

\section*{S6.\quad Why per-class evidences are not reported}
\label{sec:methods-classBF}
\subsection*{On per-class evidence}

One might expect the relative support for each family to be quantifiable as a
model-selection statistic -- a Bayes factor between classes.~A Bayes factor is a ratio of marginal likelihoods, $B_{ij}=p(\mathcal D\mid\mathcal M_i)/p(\mathcal D\mid\mathcal M_j)$, comparing two generative models $\mathcal M_i$ and $\mathcal M_j$, each equipped with a likelihood for the data $\mathcal D$ and a prior over its own parameters~\cite{Kass:1995loi}.~Our pipeline does not admit this construction.~We train a \emph{single} diffusion model on the union of the ten families, so the ``classes'' are labels on training curves rather than competing generative models:~no per-class likelihood exists, and hence no per-class marginal from which a ratio could be formed.~Retraining ten separate per-class diffusion models would not repair this.~The joint training is precisely what defines the broad prior on which the entire analysis rests; ten independently trained models would each carry its own, much narrower prior, and the resulting ratios would compare ten different priors rather than ten physical hypotheses under a common one.~Capacity is a secondary but real concern:~each per-class model would be trained on an order of magnitude less data, forcing a different architecture and breaking the like-for-like comparison.

\subsection*{Distributional similarity}

Distributional similarity is not a substitute for the missing evidences.~It is tempting to replace the marginal-likelihood ratio by a distributional similarity between the reconstruction and each training family.~Let $D$ denote the reconstructed ensemble of $c_s^2$ curves under $\chi$EFT\,+\,astro\,+\,pQCD conditioning and $C_k$ the training curves of family $k=1,\dots,10$.~Each 200-point curve is reduced to a low-dimensional feature vector -- with components such as the height and location of the $c_s^2$ maximum above $2n_0$, its value at $8n_0$, its minimum over $[1.5,6]\,n_0$, and the conformality measure at $8n_0$ -- and the feature densities $g_D$ and $g_k$ are estimated by Gaussian kernel density estimation.~Resemblance to family $k$ can then be quantified by
\begin{equation}
\mathrm{KL}\!\left(g_D\,\Vert\,g_k\right)
=\int g_D(u)\,\ln\!\frac{g_D(u)}{g_k(u)}\,\mathrm du .
\label{eq:KLrank}
\end{equation}
Read as a statement of class importance, however, any such ranking is structurally misleading, because a single-number comparison between a broad ensemble and an individual family conflates \emph{location} with \emph{breadth}.~The reconstruction is, by construction, a mixture of many admissible behaviours, and a family may contribute substantially to that mixture and yet -- being concentrated on its characteristic shapes -- differ markedly from the ensemble \emph{as a whole distribution}.

\section*{S7.\quad Supplementary tables and figures}
\label{sec:si-tabsfigs}

\newpage
\renewcommand{\figurename}{Extended Data Figure}
\renewcommand{\tablename}{Extended Data Table}


\begin{table*}[t]
\centering
\small
\renewcommand{\arraystretch}{1.2}
\setlength{\tabcolsep}{5pt}
\caption{\textbf{Posterior fiducial quantities}.~The diffusion reconstruction, compared with the Annala~2023 GP and C4 posteriors (see Sec.~\nameref{sec:si-external}).~Medians with $68\%$ credible intervals. Radii in km, $L$ in MeV, $M_{\mathrm{TOV}}$ in $M_\odot$.~$\mathcal{P}^{\mathrm{prior}}_{\mathrm{PT}}$ and $\mathcal{P}^{\mathrm{post}}_{\mathrm{PT}}$ are the prior and posterior frequencies of the necessary phase-transition signature, and $\mathrm{BF}$ is their odds ratio.~A dash indicates a quantity not extractable from the corresponding public release.~$c_{s,\max}^2$ and $n_{\mathrm{peak}}/\nsat$ are the height and density of each equation of state's \emph{own} $c_s^2(n_B)$ peak over the full density grid;~$\mathcal{P}(n_{\mathrm{peak}}\!>\!n_{\mathrm{TOV}})$ is the posterior weight for which the peak lies above the central density of the maximum-mass star.}
\label{tab:fiducials}

\begin{minipage}[t]{0.60\textwidth}
\centering
\begin{tabular}{lccc}
\hline\hline
Quantity                 & This work               & Annala GP               & Annala C4               \\
\hline
$R(1.4\,M_\odot)$        & $12.56^{+0.36}_{-0.45}$ & $12.24^{+0.18}_{-0.22}$ & $12.24^{+0.20}_{-0.20}$ \\
$R(1.6\,M_\odot)$        & $12.45^{+0.36}_{-0.44}$ & $12.29^{+0.19}_{-0.23}$ & $12.27^{+0.19}_{-0.21}$ \\
$R(2.08\,M_\odot)$       & $11.76^{+0.47}_{-0.51}$ & $12.20^{+0.27}_{-0.37}$ & $12.16^{+0.30}_{-0.40}$ \\
$\Lambda(1.4\,M_\odot)$  & $469^{+97}_{-100}$ & $465^{+42}_{-68}$       & --                      \\
$\Lambda(1.6\,M_\odot)$  & $193^{+42}_{-42}$ & $205^{+28}_{-31}$       & --                      \\
$\Lambda(2.08\,M_\odot)$ & $18.9^{+8.2}_{-6.2}$ & $30.4^{+7.4}_{-7.9}$    & --                      \\
$M_{\mathrm{TOV}}$       & $2.06^{+0.10}_{-0.08}$  & $2.23^{+0.14}_{-0.11}$  & --                      \\
$c_{s,\max}^2$           & $0.55^{+0.23}_{-0.12}$  & $0.66^{+0.13}_{-0.10}$  & $0.80^{+0.12}_{-0.16}$  \\
$n_{\mathrm{peak}}/\nsat$& $6.12^{+1.81}_{-1.96}$  & $3.48^{+2.86}_{-0.57}$  & $3.18^{+0.75}_{-0.53}$  \\
\hline\hline
\end{tabular}
\end{minipage}
\hfill
\begin{minipage}[t]{0.38\textwidth}
\centering
\begin{tabular}{lc}
\hline\hline
Quantity                      & This work               \\
\hline
$n_{\mathrm{TOV}}/\nsat$      & $6.38^{+0.47}_{-0.45}$  \\
$\cs(\mathrm{TOV})$           & $0.47^{+0.21}_{-0.08}$  \\
$\Delta(\mathrm{TOV})$        & $-0.01^{+0.04}_{-0.09}$  \\
$\gamma(\mathrm{TOV})$        & $1.39^{+0.20}_{-0.10}$  \\
$d_c(\mathrm{TOV})$           & $0.13^{+0.14}_{-0.03}$  \\
$L$                           & $60.5^{+7.5}_{-7.6}$    \\
$\mathcal{P}^{\text{post}}_{\mathrm{PT}}$   & $0.014$                 \\
$\mathcal{P}^{\text{prior}}_{\mathrm{PT}}$   & $0.088$                 \\
$\mathrm{BF}$                 & $0.14^{+0.02}_{-0.02}$  \\
$\mathcal{P}(n_{\mathrm{peak}}\!>\!n_{\mathrm{TOV}})$ & $0.451$  \\
\hline\hline
\end{tabular}
\end{minipage}
\end{table*}

\begin{table*}[t]
\centering
\small
\renewcommand{\arraystretch}{1.2}
\setlength{\tabcolsep}{5pt}
\caption{\textbf{Effect of alternative data selections on the canonical-star and phase-transition quantities.}~Entries are medians with $68\%$ credible intervals evaluated at $10^{5}$ posterior draws. Radii in km, tidal deformabilities dimensionless, $M_{\mathrm{TOV}}$ in $M_\odot$.~Baseline is the standard analysis;~the two data variants swap the PSR~J0437$-$4715 mass--radius posterior for the Miller~\textit{et al.} analysis~\cite{Miller:2025qfq} and add the maximum-mass threshold of PSR~J0952$-$0607~\cite{Romani:2022jhd} to that of PSR~J1614$-$2230.~$\mathcal{P}^{\mathrm{prior}}_{\mathrm{PT}}$ and $\mathcal{P}^{\mathrm{post}}_{\mathrm{PT}}$ are the prior and posterior frequencies of the necessary phase-transition signature, and $\mathrm{BF}$ their odds ratio.}
\label{tab:data_variants_a}
\begin{tabular}{lccc}
\hline\hline
Quantity & Baseline & J0437 (Miller) & $M_{\max}$ (J0952) \\
\hline
$R(1.4\,M_\odot)$        & $12.56^{+0.36}_{-0.45}$ & $12.80^{+0.30}_{-0.38}$ & $12.51^{+0.41}_{-0.46}$ \\
$R(1.6\,M_\odot)$        & $12.45^{+0.36}_{-0.44}$ & $12.69^{+0.32}_{-0.37}$ & $12.44^{+0.40}_{-0.45}$ \\
$R(2.08\,M_\odot)$       & $11.76^{+0.47}_{-0.51}$ & $11.98^{+0.44}_{-0.49}$ & $11.85^{+0.48}_{-0.50}$ \\
$\Lambda(1.4\,M_\odot)$  & $469^{+97}_{-100}$ & $535^{+78}_{-94}$ & $462^{+107}_{-97}$ \\
$\Lambda(1.6\,M_\odot)$  & $193^{+42}_{-42}$ & $218^{+41}_{-39}$ & $196^{+45}_{-44}$ \\
$\Lambda(2.08\,M_\odot)$ & $18.9^{+8.2}_{-6.2}$ & $21.5^{+8.3}_{-6.7}$ & $20.9^{+8.9}_{-7.0}$ \\
$M_{\mathrm{TOV}}$       & $2.06^{+0.10}_{-0.08}$  & $2.06^{+0.09}_{-0.07}$ & $2.13^{+0.10}_{-0.09}$ \\
$\mathcal{P}^{\mathrm{post}}_{\mathrm{PT}}$  & $0.014$ & $0.012$ & $0.014$ \\
$\mathcal{P}^{\mathrm{prior}}_{\mathrm{PT}}$ & $0.088$ & $0.088$ & $0.088$ \\
$\mathrm{BF}$            & $0.14^{+0.02}_{-0.02}$  & $0.13^{+0.02}_{-0.02}$ & $0.14^{+0.03}_{-0.03}$ \\
\hline\hline
\end{tabular}
\end{table*}

\begin{table*}[t]
\centering
\small
\renewcommand{\arraystretch}{1.2}
\setlength{\tabcolsep}{5pt}
\caption{\textbf{Effect of alternative data selections on the maximum-mass-configuration and peak quantities.}~Columns as in Table~\ref{tab:data_variants_a}.~Quantities labelled TOV are
evaluated at the central density of the maximum-mass star; $L$ in MeV, densities in $\nsat$. $c_{s,\max}^2$ and $n_{\mathrm{peak}}/\nsat$ are the height and density of each equation of state's \emph{own} $c_s^2(n_B)$ peak over the full density grid;~$\mathcal{P}(n_{\mathrm{peak}}\!>\!n_{\mathrm{TOV}})$ is defined as in Table~\ref{tab:fiducials}.}
\label{tab:data_variants_b}
\begin{tabular}{lccc}
\hline\hline
Quantity & Baseline & J0437 (Miller) & $M_{\max}$ (J0952) \\
\hline
$n_{\mathrm{TOV}}/\nsat$ & $6.38^{+0.47}_{-0.45}$ & $6.22^{+0.44}_{-0.40}$ & $6.23^{+0.49}_{-0.48}$ \\
$\cs(\mathrm{TOV})$      & $0.47^{+0.21}_{-0.08}$ & $0.45^{+0.18}_{-0.06}$ & $0.52^{+0.20}_{-0.11}$ \\
$\Delta(\mathrm{TOV})$   & $-0.01^{+0.04}_{-0.09}$ & $0.01^{+0.03}_{-0.08}$ & $-0.04^{+0.06}_{-0.08}$ \\
$\gamma(\mathrm{TOV})$   & $1.39^{+0.20}_{-0.10}$ & $1.38^{+0.18}_{-0.06}$ & $1.40^{+0.20}_{-0.14}$ \\
$d_c(\mathrm{TOV})$      & $0.13^{+0.14}_{-0.03}$ & $0.13^{+0.11}_{-0.02}$ & $0.16^{+0.13}_{-0.05}$ \\
$L$                      & $60.5^{+7.5}_{-7.6}$ & $64.7^{+7.4}_{-7.4}$ & $59.4^{+7.8}_{-7.9}$ \\
$c_{s,\max}^2$           & $0.55^{+0.23}_{-0.12}$ & $0.48^{+0.25}_{-0.07}$ & $0.66^{+0.20}_{-0.19}$ \\
$n_{\mathrm{peak}}/\nsat$& $6.12^{+1.81}_{-1.96}$ & $6.08^{+1.85}_{-1.39}$ & $5.74^{+2.22}_{-2.04}$ \\
$\mathcal{P}(n_{\mathrm{peak}}\!>\!n_{\mathrm{TOV}})$ & $0.451$ & $0.462$ & $0.417$ \\
$N_{\mathrm{eff}}$ & $4187.3$ & $5790.3$ & $2113.9$ \\
\hline\hline
\end{tabular}
\end{table*}

\begin{table*}[t]
\centering
\small
\renewcommand{\arraystretch}{1.2}
\setlength{\tabcolsep}{5pt}
\caption{\textbf{Robustness of the canonical-star and phase-transition quantities.}
Each column repeats the full inference at a matched sample size of $10^{5}$ draws;~entries are medians with $68\%$ credible intervals.~Radii in km, tidal deformabilities dimensionless.~Columns $M_{\mathrm{I}}$, $M_{\mathrm{II}}$ and $M_{\mathrm{III}}$ add three parametrized families to the training ensemble;~$\Lambda=450$ rebuilds the $\chi$EFT anchors at the lower cutoff.~$\mathcal{P}^{\mathrm{prior}}_{\mathrm{PT}}$ and $\mathcal{P}^{\mathrm{post}}_{\mathrm{PT}}$ are the prior and posterior frequencies of the necessary phase-transition signature, and $\mathrm{BF}$ their odds ratio.}
\label{tab:robustness_a}
\begin{tabular}{lccccc}
\hline\hline
Quantity & Baseline & $M_{\mathrm{I}}$ & $M_{\mathrm{II}}$ & $M_{\mathrm{III}}$ & $\Lambda\!=\!450$ \\
\hline
$R(1.4\,M_\odot)$ & $12.56^{+0.36}_{-0.45}$ & $12.58^{+0.35}_{-0.45}$ & $12.50^{+0.39}_{-0.46}$ & $12.54^{+0.37}_{-0.45}$ & $12.43^{+0.42}_{-0.46}$ \\
$R(1.6\,M_\odot)$ & $12.45^{+0.36}_{-0.44}$ & $12.45^{+0.36}_{-0.44}$ & $12.38^{+0.39}_{-0.43}$ & $12.44^{+0.36}_{-0.43}$ & $12.32^{+0.40}_{-0.44}$ \\
$R(2.08\,M_\odot)$ & $11.76^{+0.47}_{-0.51}$ & $11.75^{+0.45}_{-0.48}$ & $11.71^{+0.47}_{-0.48}$ & $11.84^{+0.51}_{-0.52}$ & $11.69^{+0.56}_{-0.49}$ \\
$\Lambda(1.4\,M_\odot)$ & $469^{+97}_{-100}$ & $469^{+97}_{-98}$ & $454^{+103}_{-95}$ & $463^{+99}_{-99}$ & $425^{+99}_{-95}$ \\
$\Lambda(1.6\,M_\odot)$ & $193^{+42}_{-42}$ & $193^{+43}_{-42}$ & $186^{+44}_{-40}$ & $191^{+42}_{-41}$ & $175^{+44}_{-41}$ \\
$\Lambda(2.08\,M_\odot)$ & $18.9^{+8.2}_{-6.2}$ & $18.5^{+8.1}_{-5.8}$ & $18.4^{+8.4}_{-5.9}$ & $20.1^{+9.7}_{-6.6}$ & $18.1^{+9.2}_{-6.0}$ \\
$M_{\mathrm{TOV}}$ & $2.06^{+0.10}_{-0.08}$ & $2.06^{+0.09}_{-0.08}$ & $2.06^{+0.11}_{-0.08}$ & $2.07^{+0.10}_{-0.08}$ & $2.07^{+0.10}_{-0.08}$ \\
$\mathcal{P}^{\mathrm{post}}_{\mathrm{PT}}$ & $0.014$ & $0.014$ & $0.015$ & $0.005$ & $0.007$ \\
$\mathcal{P}^{\mathrm{prior}}_{\mathrm{PT}}$ & $0.088$ & $0.081$ & $0.060$ & $0.111$ & $0.086$ \\
$\mathrm{BF}$ & $0.14^{+0.02}_{-0.02}$ & $0.16^{+0.03}_{-0.03}$ & $0.24^{+0.04}_{-0.04}$ & $0.04^{+0.01}_{-0.01}$ & $0.08^{+0.03}_{-0.02}$ \\
\hline\hline
\end{tabular}
\end{table*}

\begin{table*}[t]
\centering
\small
\renewcommand{\arraystretch}{1.2}
\setlength{\tabcolsep}{5pt}
\caption{\textbf{Robustness of the maximum-mass-configuration and peak quantities.} Columns as in Table~\ref{tab:robustness_a}.~Quantities labelled TOV are evaluated at the central density of the maximum-mass star;~$L$ in MeV, densities in $\nsat$.~$c_{s,\max}^2$ and $n_{\mathrm{peak}}/\nsat$ are the height and density of each equation of state's \emph{own} $c_s^2(n_B)$ peak over the full density grid;~$\mathcal{P}(n_{\mathrm{peak}}\!>\!n_{\mathrm{TOV}})$ is defined as in Table~\ref{tab:fiducials}.~Matched sample size is $10^{5}$ draws.}
\label{tab:robustness_b}
\begin{tabular}{lccccc}
\hline\hline
Quantity & Baseline & $M_{\mathrm{I}}$ & $M_{\mathrm{II}}$ & $M_{\mathrm{III}}$ & $\Lambda\!=\!450$ \\
\hline
$n_{\mathrm{TOV}}/\nsat$ & $6.38^{+0.47}_{-0.45}$ & $6.41^{+0.49}_{-0.45}$ & $6.40^{+0.50}_{-0.46}$ & $6.28^{+0.48}_{-0.50}$ & $6.42^{+0.48}_{-0.52}$ \\
$\cs(\mathrm{TOV})$ & $0.47^{+0.21}_{-0.08}$ & $0.48^{+0.22}_{-0.10}$ & $0.50^{+0.21}_{-0.11}$ & $0.45^{+0.20}_{-0.09}$ & $0.49^{+0.19}_{-0.11}$ \\
$\Delta(\mathrm{TOV})$ & $-0.01^{+0.04}_{-0.09}$ & $-0.02^{+0.05}_{-0.09}$ & $-0.02^{+0.06}_{-0.09}$ & $-0.00^{+0.04}_{-0.08}$ & $-0.02^{+0.06}_{-0.08}$ \\
$\gamma(\mathrm{TOV})$ & $1.39^{+0.20}_{-0.10}$ & $1.40^{+0.21}_{-0.13}$ & $1.41^{+0.22}_{-0.12}$ & $1.36^{+0.22}_{-0.19}$ & $1.39^{+0.20}_{-0.17}$ \\
$d_c(\mathrm{TOV})$ & $0.13^{+0.14}_{-0.03}$ & $0.14^{+0.14}_{-0.04}$ & $0.15^{+0.14}_{-0.05}$ & $0.12^{+0.13}_{-0.04}$ & $0.14^{+0.13}_{-0.05}$ \\
$L$ & $60.5^{+7.5}_{-7.6}$ & $61.3^{+7.3}_{-7.8}$ & $60.5^{+7.5}_{-8.2}$ & $61.5^{+7.3}_{-8.3}$ & $66.8^{+7.5}_{-7.8}$ \\
$c_{s,\max}^2$ & $0.55^{+0.23}_{-0.12}$ & $0.58^{+0.20}_{-0.15}$ & $0.63^{+0.19}_{-0.20}$ & $0.64^{+0.18}_{-0.20}$ & $0.64^{+0.18}_{-0.18}$ \\
$n_{\mathrm{peak}}/\nsat$ & $6.12^{+1.81}_{-1.96}$ & $6.34^{+1.62}_{-2.15}$ & $6.53^{+1.43}_{-2.49}$ & $5.40^{+2.53}_{-2.15}$ & $5.70^{+2.26}_{-2.30}$ \\
$\mathcal{P}(n_{\mathrm{peak}}\!>\!n_{\mathrm{TOV}})$ & $0.451$ & $0.498$ & $0.527$ & $0.377$ & $0.401$ \\
$N_{\mathrm{eff}}$ & $4187.3$ & $4212.2$ & $4349.2$ & $3387.0$ & $1714.7$ \\
\hline\hline
\end{tabular}
\end{table*}

\begin{table*}[t]
\centering
\small
\renewcommand{\arraystretch}{1.2}
\setlength{\tabcolsep}{5pt}
\caption{\textbf{$\beta$-equilibrium sound-speed anchors.} Monte-Carlo
means and $1\sigma$ uncertainties from the joint PNM--SNM draw at
N$^3$LO ($\Lambda_b=600\,\mathrm{MeV}$, $2\times10^4$ samples), at the
five densities used for conditioning. The two rows correspond to the
chiral-interaction regulator cutoffs $\Lambda=500$ and $450$~MeV.}
\label{tab:anchors}
\begin{tabular}{lccccc}
\hline\hline
$\nB/\nsat$ & $0.50$ & $0.76$ & $0.99$ & $1.25$ & $1.52$ \\
\hline
$10^2\,\cs$ \ ($\Lambda=500$~MeV) & $1.11\pm0.16$ & $2.81\pm0.45$ & $4.88\pm0.85$ & $7.68\pm1.55$ & $10.29\pm2.54$ \\
$10^2\,\cs$ \ ($\Lambda=450$~MeV) & $1.30\pm0.14$ & $3.08\pm0.40$ & $4.96\pm0.75$ & $7.52\pm1.38$ & $9.61\pm2.28$ \\
\hline\hline
\end{tabular}
\end{table*}
\clearpage
\begin{figure}[h]
  \begin{center}
    \vspace{-1cm}
\includegraphics[width=0.65\textwidth,keepaspectratio,angle=0,clip]{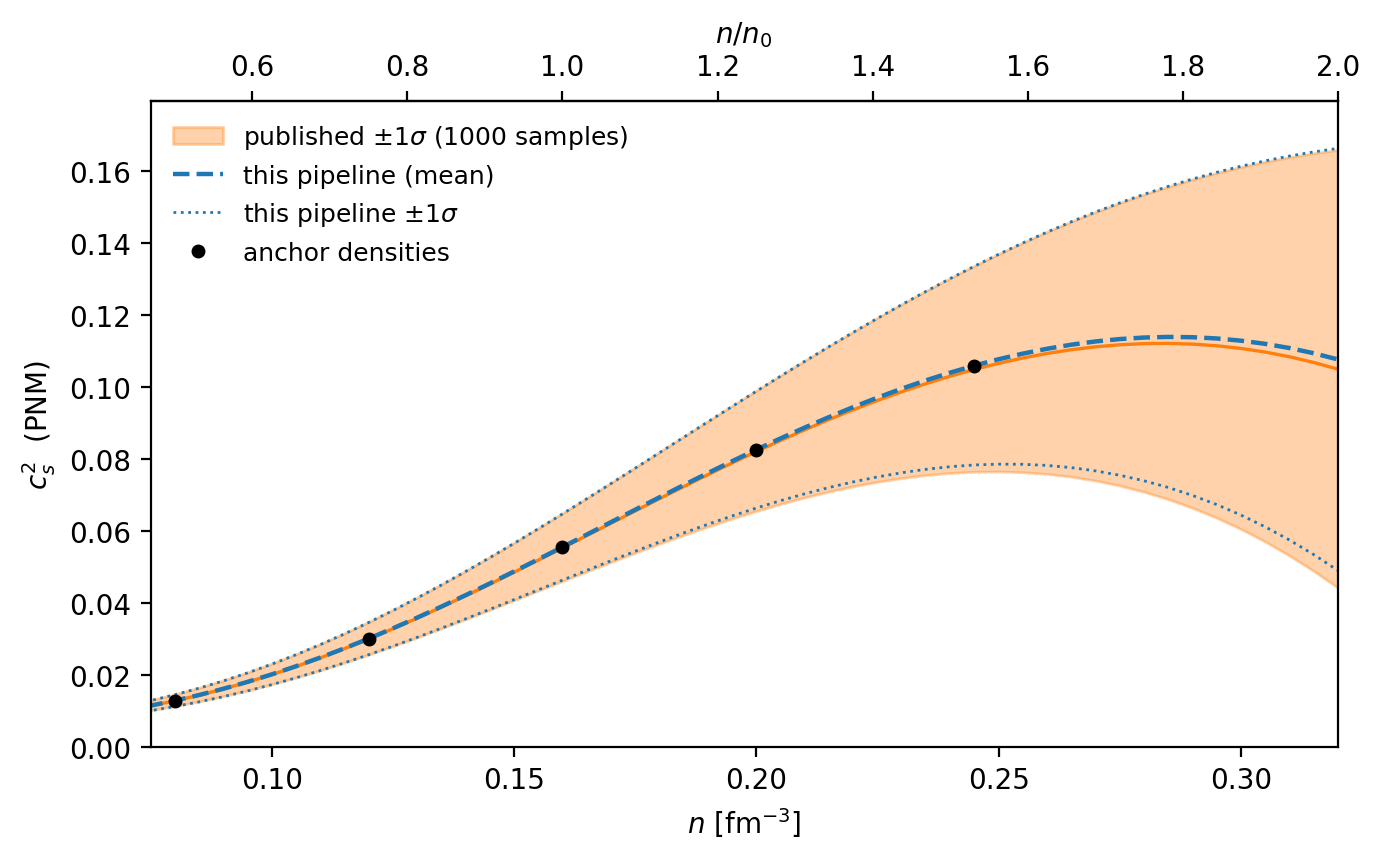}
\end{center}
\linespread{1.3}\selectfont{}
\caption{
\textbf{PNM-mode validation of the sampling pipeline.}~Squared sound speed of pure neutron matter at N$^3$LO ($\Lambda=500$~MeV, $\Lambda_b=600$~MeV).~Orange: mean and $\pm1\sigma$ band of the 1,000 Monte-Carlo samples published by the BUQEYE Collaboration~\cite{Drischler:2020hwi,BUQEYE_NM_repo};~blue dashed and dotted curves: mean and $\pm1\sigma$ from this work, run in PNM-only mode.~Black circles mark the five anchor densities.~At the anchor densities the means agree to  $1\%$ and the widths to $3.5\%$, consistent with the ${\approx}2\%$ finite-sample precision of the published sample set;~the visible
deviation of the lower band edge lies beyond the anchor region.
\label{fig:BEQ-our}
}
\end{figure}

\begin{figure}[t]
  \begin{center}
    \vspace{-1cm}
\includegraphics[width=0.65\textwidth,keepaspectratio,angle=0,clip]{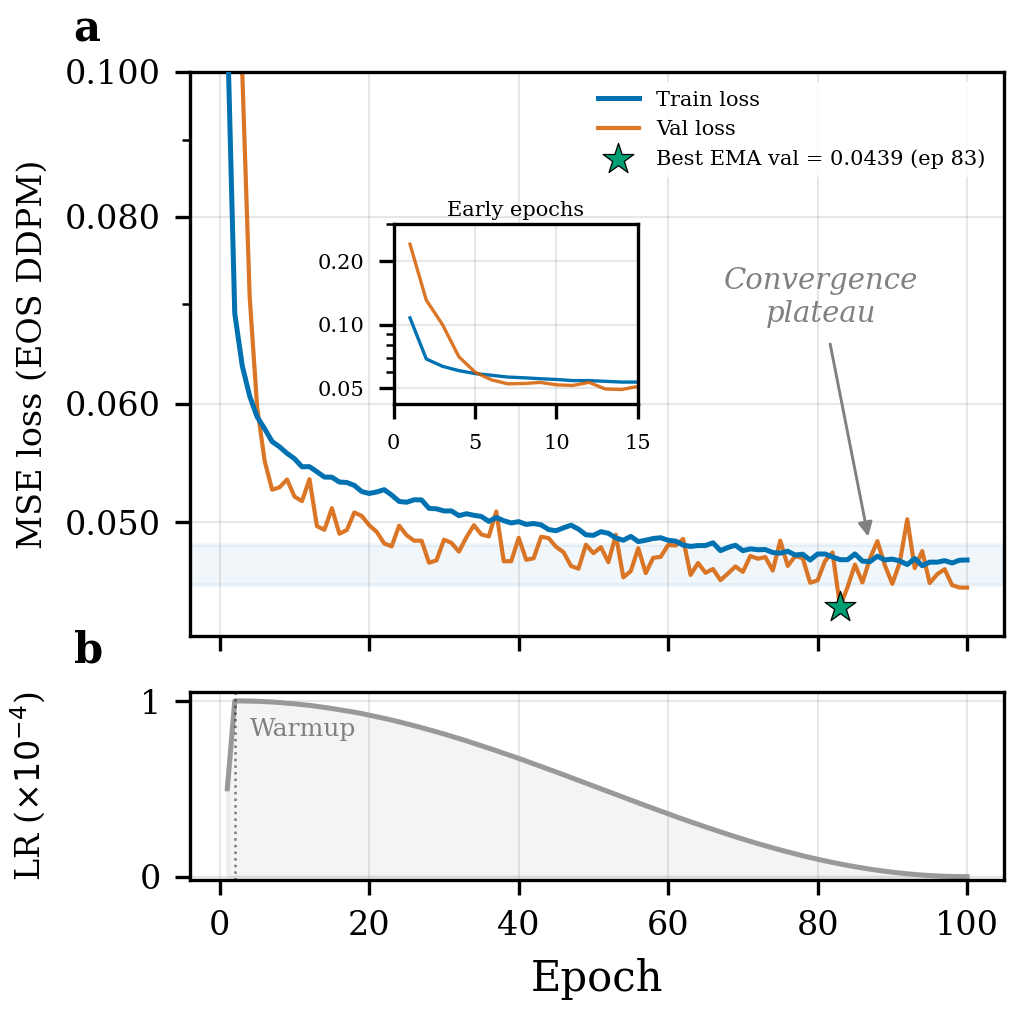}
\end{center}
\linespread{1.3}\selectfont{}
\caption{ \textbf{Training dynamics of the equation-of-state diffusion model.}~\textbf{a,} Mean squared error (MSE) loss on the training (blue) and validation (orange) sets over 100 epochs, on a logarithmic scale.~The shaded band spans the 10th--90th percentiles of the validation loss over the final 40 epochs;~the star marks the lowest validation loss attained by the exponential moving average (EMA) of the weights, $0.0439$ at epoch 83.~Inset, the first 15 epochs at full vertical scale.~\textbf{b}, Learning-rate schedule: linear warm-up over two epochs to a peak of $1\times10^{-4}$, followed by cosine decay to zero.~The model ($19.4\times10^{6}$ parameters) was trained on $10^{6}$ equation-of-state curves sampled from ten model classes, with $3\times10^{4}$ curves held out for validation.
\label{fig:plateau}
}
\end{figure}

\end{document}